\documentclass[%
 reprint,
bibnotes,
 amsmath,amssymb,
 aps,
letterpaper
]{revtex4-2}

\usepackage{graphicx}
\usepackage{dcolumn}
\usepackage{bm}
\usepackage[dvipsnames]{xcolor}
\usepackage[normalem]{ulem}
\usepackage{hyperref}

\def\Nup{N_\uparrow}
\def\Ndn{N_\downarrow}

\def\kup{\left\vert\uparrow\right\rangle}
\def\kdn{\left\vert\downarrow\right\rangle}
\def\Tup{T_\uparrow}
\def\Tdn{T_\downarrow}
\def\omex{\omega_\mathrm{ex}}

\def\aupdn{a_{\uparrow\downarrow}}
\def\dd{\mathrm{d}}
\newcommand{\ket}[1]{\left\vert #1 \right\rangle}

\newcommand{\avg}[1]{\langle #1 \rangle}

\DeclareMathOperator{\Var}{Var}
\let\rm=\mathrm
\let\bf=\mathbf

\begin{document}

\title{Observing spin-squeezed states under spin-exchange collisions for a second}

\author{Meng-Zi Huang$^1$}
\email{Present address: Institute for quantum electronics, ETH Zurich, Switzerland}
\author{Jose Alberto de la Paz$^2$}
\author{Tommaso Mazzoni$^2$}
\email{Present address: WeLinQ SAS, Paris, France}
\author{Konstantin~Ott$^{1}$}
\email{Present address: VITRONIC Dr.-Ing. Stein Bildverarbeitungssysteme GmbH, Wiesbaden,
Germany}
\author{Peter Rosenbusch$^{2}$}
\email{Present address: Muquans/Exail, Talence, France}
\author{Alice Sinatra$^1$}
\author{Carlos L. Garrido Alzar$^2$}
\author{Jakob Reichel$^1$}
\email{jakob.reichel@ens.fr}
\affiliation{
$^1$Laboratoire Kastler Brossel, ENS-Universit{\'e} PSL, CNRS, Sorbonne Universit\'e, Coll{\`e}ge de France, 24 rue Lhomond, 75005 Paris, France\\
$^2$LNE-SYRTE, Observatoire de Paris-Universit{\'e} PSL, CNRS, Sorbonne Universit{\'e}, 61 Avenue de l'Observatoire, 75014 Paris, France
}
\date{May 08, 2023}

\begin{abstract}
Using the platform of a trapped-atom clock on a chip, we observe the time evolution of spin-squeezed hyperfine clock states in ultracold rubidium atoms on previously inaccessible timescales up to 1\,s.
The spin degree-of-freedom remains squeezed after 0.6\,s, which is consistent with the limit imposed by particle loss and is compatible with typical Ramsey times in state-of-the-art microwave clocks.
The results also reveal a surprising spin-exchange interaction effect that amplifies the cavity-based spin measurement via a correlation between spin and external degrees of freedom.
These results open up perspectives for squeezing-enhanced atomic clocks in a metrologically relevant regime and highlight the importance of spin interactions in real-life applications of spin squeezing.

\end{abstract}

\maketitle
\section{Introduction}
\label{sec:intro}
Spin squeezing in atomic ensembles \cite{Kitagawa1993,Wineland1994,Ma2011,Pezze2018} is a fascinating manifestation of many-particle entanglement as well as one of the most promising quantum technologies. By using entanglement to reduce the quantum projection noise in a collection of atomic spins, spin squeezing removes a limit that is already present in state-of-the-art atomic fountain clocks \cite{Santarelli1999}, inertial sensors \cite{Gauguet2009,Rosi2014}, optical lattice clocks \cite{Oelker2019} and magnetometers \cite{Wasilewski2010}. Groundbreaking experiments have demonstrated methods for creating spin-squeezed states \cite{Appel2009,Leroux2010,Schleier-Smith2010a,Riedel2010,Gross2010,Bohnet2014,Cox2016,Bohnet2016,Hosten2016}
and proof-of-principle clocks and magnetometers have been realized, with special emphasis on alkali atoms such as rubidium because they are used in the vast majority of atomic metrology devices \cite{Leroux2010a,Hosten2016,Sewell2012,Bao2020}, and recently also on optical transitions 
\cite{Pedrozo2020,robinson_direct_2022}.
Squeezing up to 20\,dB \cite{Hosten2016} has been achieved,
while even a more modest reduction would be sufficient to make quantum projection noise negligible in existing atomic clocks and sensors. However, previous squeezing experiments with alkali atoms have been limited to time scales of a few milliseconds (e.g.~5\,ms in Ref.~\cite{Leroux2010a}, 2\,ms in Ref.~\cite{Cox2016}, 1\,ms in Ref.~\cite{Hosten2016}, and 8\,ms in Ref.~\cite{Malia2020}),
while interrogation times in real clocks and sensors are typically 10--100 times longer \cite{Guena2012,Ludlow2015,Barrett2016}. How squeezed states evolve on these time scales is a question that experiments have not yet been able to address due to technical limits, masking the more fundamental effects that can be expected to arise from atomic interactions. In particular, collision-induced spin interactions are known to play an important role both in microwave \cite{Weyers2018,Deutsch2010} and optical \cite{Ludlow2015} atomic clocks. For Rb at densities typical of spin-squeezing experiments, the spin exchange rate $\omex/2\pi=2\hbar|\aupdn|\bar{n}/m$ (where $\aupdn$ is the scattering length between clock states, $\hbar$ is the reduced Planck constant, $\bar{n}$ is the atomic density, and $m$ is the atomic mass) is on the order of a few hertz, so that its effects can indeed be expected on the unexplored but relevant time scales above 100\,ms.
The role of these interactions for spin squeezing poses considerable challenges for theoretical models and is only starting to be explored \cite{Martin2013,He2019,bilitewski_dynamical_2021}, especially in realistic systems where inhomogeneities are present both for internal and external degrees of freedom.
Studying the time evolution of such interacting many-body system further enriched by the entangled spin-squeezed states is of fundamental and practical interest for quantum metrology, both for optical lattice clocks at the frontier of precision and for alkali-atom sensors which are the workhorse for atomic metrology in a broader sense.

Here, we investigate measurement-based spin squeezing in an optical cavity \cite{Schleier-Smith2010a} in the platform of a trapped-atom clock on a chip \cite{Deutsch2010,Szmuk2015},
where the coherence lifetime exceeds 20\,s.
Starting with a spin-squeezed state with up to 8.6\,dB of metrological squeezing, we measure its evolution over 1\,s in conditions typical of a metrology-grade experiment and observe the effect of spin-exchange interactions, which manifests itself in a correlation between spin and external degrees of freedom due to the cavity interaction. Their interplay gives rise to a new and surprising feedback mechanism that can amplify the cavity measurement.
Similar manifestations of spin interactions are likely to be observed in other experiments as their coherence times increase toward metrologically useful values. Taking the interaction effect into account, we can nevertheless infer that the metrological squeezing is preserved for 0.6\,s, consistent with the fundamental limit imposed by particle loss in our system.
These results are an important step on the way to squeezing-enhanced clocks and sensors with metrologically relevant stability and they highlight the importance of spin interactions in the regime of long interaction time that these instruments require.

\begin{figure}[htb]
\includegraphics[width=0.48\textwidth]{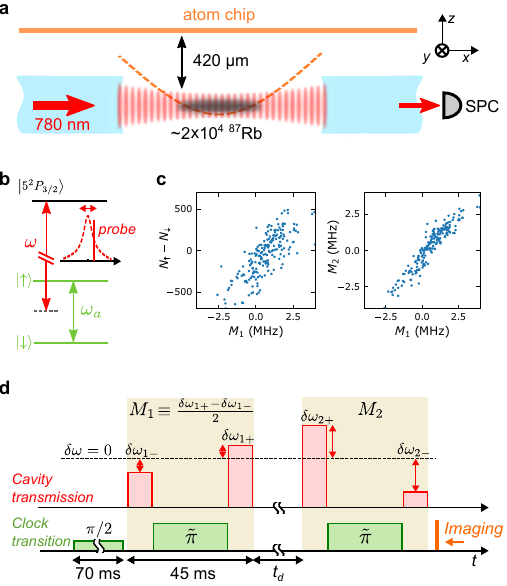}
\caption{\label{fig:scheme} {The concept of the experiment.} \textbf{a}, The experimental setup. $^{87}$Rb atoms are magnetically trapped by the atom chip (the trap potential is illustrated by the dashed orange curve). Transmitted photons are collected by a single-photon counter (SPC). Absorption images are taken along $\hat{y}$. Cavity-locking light at 1560\,nm is not shown.
  \textbf{b}, The energy level structure and the cavity-probing scheme. \textbf{c}, Typical data of the correlation of a cavity measurement with absorption imaging (left) and two consecutive cavity measurements, $M_1$ and $M_2$ (right). The correlation with absorption imaging is limited by imaging noise, $\Delta(\Nup-\Ndn)\sim 100$, comparable to the SQL.
\textbf{d}, An experimental sequence with two composite cavity measurements, $M_1$ and $M_2$, for squeezing verification.
The green boxes represent pulses on the clock transition and the red boxes cavity-probe transmission, from which $\delta\omega_\pm$ are deduced. The delay $t_d$ between measurements varies from a few milliseconds to 1\,s. $\tilde{\pi}$ denotes a composite $\pi$ pulse.
}
\end{figure}

\section{Experimental setup}\label{sec:exp}

Our experiment is similar to the trapped-atom clock on a chip (TACC) described in Refs.~\cite{Deutsch2010,Szmuk2015} but additionally contains a fiber Fabry-P{\'erot} cavity \cite{Ott2016}. An ensemble of $N\sim2\times10^4$ $^{87}$Rb atoms is magnetically trapped inside this cavity using an atom chip [Fig.~\ref{fig:scheme}(a)]. The trap is cigar shaped, with frequencies $\lbrace \omega_x, \omega_y, \omega_z \rbrace/2\pi \approx \lbrace 7.5, 122, 113 \rbrace\,$Hz, with the cavity axis along $\hat{x}$. At typical temperatures of $T\approx 200\,$nK transversely, the cloud is in the collisionless regime such that each atom preserves its motional energy over many oscillations in the trap (see the time scales in Appendix~\ref{app:timescales}).

The hyperfine states $\kdn \equiv \ket{F=1,m_F=-1}$ and $\kup \equiv \ket{F=2,m_F=1}$ are chosen as clock states \cite{Harber2002,Deutsch2010}. Used as a clock with standard Ramsey interrogation and coherent spin states (CSSs), the experiment currently reaches a fractional frequency stability of $6.5\times 10^{-13}\,\rm{s}^{-1/2}$ and has a phase-coherence time on the order of 20\,s, longer than the trap lifetime due to background loss.

 We consider the collective spin vector
$\hat{\bf{S}}=\sum_i^N\hat{\bf{s}}_i$ with $\hat{\bf{s}}_i=\lbrace\hat{\sigma}_x^{(i)}, \hat{\sigma}_y^{(i)},\hat{\sigma}_z^{(i)}\rbrace/2$, where the $\hat{\sigma}_{x,y,z}^{(i)}$ are Pauli matrices for the $i$th atom. The measurement of the $z$ component is given by the population difference
$S_z = (\Nup-\Ndn)/2$, where $\Nup$ ($\Ndn$) is the atom number in $\kup$ ($\kdn$). For a CSS, the fluctuation in $S_z$ is given by the standard quantum limit (SQL): $\Delta^2 S_z|_\rm{CSS}=N/4$, where $\Delta^2$ denotes the standard variance.
Spin squeezing is generated by a quantum nondemolition (QND) measurement of the collective spin observable $\hat{S}_z$
via the frequency shift $\delta\omega$ that it induces to
an off-resonant optical cavity \cite{Schleier-Smith2010a}. The cavity has a mode waist ($1/e$ radius) $w_0=13.6\,\mu$m, length $L=1215(20)\,\mu$m, and line width (full width at half maximum) $\kappa/2\pi=45.8(6)\,$MHz. It is tuned midway between the 780-nm $D_2$ transitions $\kdn\rightarrow 5P_{3/2}$ and $\kup\rightarrow 5P_{3/2}$, such that to a good approximation
$\delta\omega = \Omega_e S_z$, where $\Omega_e=\sum_i^N\Omega_i/N$
is the ensemble-averaged shift per spin flip and $\Omega_i$ is the coupling strength of the $i$th atom. The value $\Omega_e= 2\pi\times 16.2(3)\,$kHz
has an uncertainty limited by the temperature and is determined experimentally by measuring the cavity shift after preparing a CSS with different $\avg{S_z}$ (Appendix \ref{app:calib}).
In the following experiments,
we measure $\delta\omega$ with a probe laser blue detuned from the cavity resonance by approximately $\kappa/2$ [Fig.~\ref{fig:scheme}(b)] and detect the transmitted photons using a single-photon counter, with an overall detection efficiency $\eta = 0.63(2)$.
Additionally, $\Nup$ and $\Ndn$ are also measured by absorption imaging after the time of flight (TOF). We verify that both measurements agree to within the noise of the absorption imaging, which is close to the SQL [Fig.~\ref{fig:scheme}(c)].

In our experiment, the inhomogeneity of the coupling $\Omega_i$ is predominantly in the transverse directions due to the cavity intensity profile, whereas it is almost averaged out by atomic motion along the cavity axis.
In order to reduce inhomogeneity-induced dephasing \cite{Schleier-Smith2010a,Bohnet2014}, we fix the probe-pulse duration to the vertical trap period, $\tau_p=8.85\,\mbox{ms}=2\pi/\omega_z\approx2\pi/\omega_y$. Thus $\Omega_i$ only depends on the transverse motional energy of an atom and remains constant until a lateral collision occurs (Appendix~\ref{app:inhomo}). The remaining inhomogeneity between atoms with different motional energies is further suppressed by employing a spin-echo sequence as in previous experiments \cite{Schleier-Smith2010a,Bohnet2014}. A complete cavity measurement is then composed of two cavity-probe pulses separated by a $\pi$ pulse on the clock transition [Fig.~\ref{fig:scheme}(d)]. The measured $S_z$ is deduced from the cavity shifts $\delta\omega_\pm$ of the two probe pulses as $S_z=M/\Omega_e$ where $M\equiv(\delta\omega_{+} - \delta\omega_{-})/2$.

\begin{figure}[tb]
\includegraphics[width=0.45\textwidth]{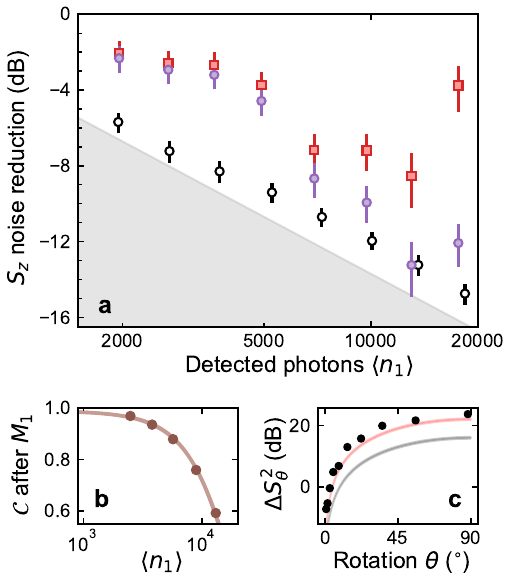}
\caption{\label{fig:sqz}
{The spin squeezing characterization.}
\textbf{a}, Conditional squeezing results for $N=1.8(1)\times10^4$ atoms and minimum delay ($t_d=6\,$ms) between measurements $M_1$ and $M_2$. The variance values are normalized to the SQL ($N/4$) and expressed in decibels. The cavity measurement without atoms (open circles) approaches the photon-shot-noise (PSN) limit (the boundary of the shaded zone). The number squeezing $\xi_N^2$ (purple circles) (Eq.~\ref{eq:cond_sz})
decreases to below $-12\,$dB.
After normalizing by the independently measured coherence (panel \textbf{b}), we obtain the metrological squeezing (red squares), reaching an optimum of $8.6^{+1.8}_{-1.3}$\,dB for $13.0(2)\times10^3$ photons. The error bars indicate $1\sigma$ confidence and are obtained with a bootstrapping method. \textbf{b}, The Ramsey contrast $\mathcal{C}$ as a function of the measurement strength. The curve is a fit to $\mathcal{C}=\exp[-\avg{n_1}/\gamma_1 -\left(\avg{n_1}/\gamma_2\right)^2]$ (Appendix~\ref{app:coherence}). Its thickness indicates the fit uncertainty. \textbf{c}, The spin-noise tomography at $\avg{n_1}=8.9(2)\times 10^3$, measured by inserting, between $M_1$ and $M_2$, a rotation $\theta$ around $\avg{\hat{\bf{S}}}$. The gray curve represents the theoretical minimum-uncertainty state, while the pink curve takes into account the
phase noise induced by the PSN of $M_1$.
}
\end{figure}

\section{Spin squeezing by QND measurement}\label{sec:squeezing}

We first investigate the metrological spin squeezing generated by our cavity-QND measurement. As shown in Fig.~\ref{fig:scheme}(d),
we start with all atoms in $\kdn$ and apply a $\pi/2$ pulse on the clock transition to prepare a CSS on the equator of the Bloch sphere. A composite cavity measurement $M_1$ measures the cavity detuning to determine $S_z$.
A second identical measurement $M_2$ after a minimum delay $t_d = 6\,$ms  serves to verify the measurement uncertainty and spin squeezing.
Noise is quantified from the variance of 200 repetitions of this sequence. First, we perform this protocol with no atoms in the cavity to determine the noise floor [Fig.~\ref{fig:sqz}(a), open black circles]. The result is close to the photon shot noise (PSN) of the detected photons, given by $\Delta^2 M_l^\rm{psn} \approx \kappa^2/(4\avg{n_{l}})$, where $\avg{n_{l}}$ is the average number of detected photons per measurement ($l=1,2$).
For the atom number $N=1.8(1)\times 10^4$ used here, the PSN falls below the SQL for $\avg{n_1}\gtrsim 1000$ detected photons, allowing for spin-noise reduction by the cavity measurement.

A QND measurement produces ``conditional squeezing'':  $M_1$ yields a different result every time, following the quantum fluctuations of the CSS. The squeezing manifests itself in the correlation with the second measurement $M_2$, which for a squeezed state agrees with $M_1$ to better than the SQL \cite{Pezze2018}.
With two measurements performed on the same sample with negligible delay, the spin noise of the state after $M_1$ can be quantified as
\begin{equation}
  \Delta^2S_z |_{M_1}=
  \Omega_e^{-2}\left[\Var\left(M_2 -
    \zeta M_1\right) - \Delta^2 M_2^\rm{psn}\right]
    \,,
    \label{eq:cond_sz}
\end{equation}
where $\zeta$ is chosen such that it minimizes the variance (and hence accounts for systematic differences between the two measurements) \cite{Appel2009}. 
To assess the spin noise after $M_1$, it is legitimate to subtract the detection noise of the verification measurement $M_2$, which contains the PSN, $\Delta^2 M_2^\rm{psn}$, plus technical noises such as cavity-lock fluctuations.
In Eq.~\ref{eq:cond_sz}, we conservatively subtract only the PSN, so that we obtain an upper bound for $\Delta^2 S_z$. Fig.~\ref{fig:sqz}(a) (purple circles) shows it as a function of the number of detected photons in $M_1$. It is normalized to the SQL to give the number squeezing $\xi_N^2=4\Delta^2 S_z/N$ \cite{Kitagawa1993}.
The metrological squeezing $\xi^2 = N\Delta^2 S_z/|\avg{\hat{\bf{S}}}|^2 = \xi_N^2/\mathcal{C}^2$,
which characterizes the enhancement in angular resolution on the Bloch sphere compared to the SQL \cite{Wineland1994}, additionally requires assessing the coherence, namely, the Ramsey-fringe contrast $\mathcal{C}=2|\avg{\hat{\bf{S}}}|/N$. We do this by applying a second $\pi/2$ pulse with a variable phase after $M_1$ and then measuring $S_z$ by imaging [Fig.~\ref{fig:sqz}(b)]. The contrast decay for increasing photon number is likely due to the imperfect light-shift cancellation in the spin echo [Fig.~\ref{fig:scheme}(d) and Appendix~\ref{app:coherence}]. The resulting Wineland squeezing factor is shown as red squares in Fig.~\ref{fig:sqz}(a).
We obtain an optimum metrological squeezing of $8.6^{+1.8}_{-1.3}$\,dB, with about 13\,000 detected probe photons. The optimum results from the competition between photon shot noise, which favors a higher photon number, and photon-induced decoherence.

To fully characterize the squeezed state, we also perform spin-noise tomography \cite{Schleier-Smith2010a,Cox2016} by inserting a pulse on the clock transition between $M_1$ and $M_2$ to rotate the noise distribution around $\avg{\bf{S}}$ [Fig.~\ref{fig:sqz}(c) and Appendix~\ref{app:tomo}]. The data with $\avg{n_1}=8.9(2)\times 10^3$ show an excess antisqueezing of 7.4\,dB (at $90^\circ$ rotation) above the minimum-uncertainty state (gray curve), mostly due to the shot-to-shot phase noise caused by the PSN in $M_1$ (predicted by the pink curve).

\begin{figure}[tb]
\includegraphics[width=0.4\textwidth]{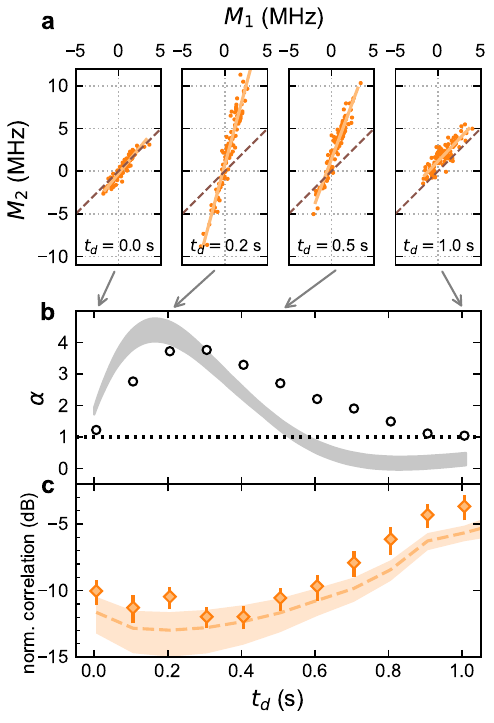}
\caption{\label{fig:corr_t}
{Amplification of the cavity measurement.}
\textbf{a}. Raw data (measured cavity detunings) of $M_2$ versus $M_1$ at $t_d=0,~0.2,~0.5,~1$\,s, respectively. The initial atom number and the average photon number are $N=2.1(1)\times10^4$ and $\avg{n_1}=9.4(2)\times10^3$ respectively. The dashed lines indicate $M_2=M_1$.
The slope of a linear fit (solid lines) gives the amplification factor in (b).
\textbf{b}. The amplification factor $\alpha$ (open circles) as a function of time $t_d$ between $M_1$ and $M_2$. The errorbars are smaller than the symbol size. The gray curve represents a semiclassical numerical simulation with no fitting (see the text and Appendix \ref{app:nsim}).
\textbf{c}. The correlation evaluated by $\Var(M_1-M_2/\alpha)$ normalized to the SQL at $t=0$. For all times, $M_2$ remains correlated with $M_1$ to below the SQL of the state prepared by $M_1$. The dashed curve is a lower bound given by our model (see the text and Appendix \ref{app:noise_model}).
}
\end{figure}

\section{Long-time evolution}\label{sec:evol}
The long phase-coherence time in our experiment allows us to observe the evolution of the spin-squeezed states over much longer time scales than in previous experiments. We do so by performing the verification measurement $M_2$ after longer times $t_d$ up to 1\,s. Tracing $M_2$ as a function of $M_1$, we find that strong linear correlation between the measurements persists for all measurement times but, surprisingly, its slope depends on $t_d$ [Fig.~\ref{fig:corr_t}(a)]. The slope $\alpha$ of a linear fit to the data increases to values up to approximately 4 for times $t_d\lesssim 300\,$ms, then decays back to values close to 1 [Fig.~\ref{fig:corr_t}(b)]. Using absorption imaging, we confirm that $S_z$ itself does not measurably evolve when $\alpha$ increases, indicating that the amplification of $M_2$ is linked to the measurement rather than to the spin state itself. We come back to the mechanism causing this amplification below. A direct way to quantify the correlation between $M_1$ and $M_2$ is to ask how much one can learn about $M_1$ from $M_2$, i.e.\ to compute $\Var(M_1-M_2/\alpha)$. The result is shown in Fig.~\ref{fig:corr_t}(c), normalized to the SQL at $t_d=0$, i.e., $\Omega_e^2 N(0)/4$. It remains 4\,dB below the SQL even for our longest measurement time of $t_d=1\,$s. 

\section{Modeling the time evolution: Losses and spin-orbit correlation}\label{sec:ampli}
The time evolution of spin-squeezed states under realistic conditions is still an open field, even theoretically. One effect that has been studied is atom loss \cite{Li2008}.  In our experiment, background gas collisions reduce the total atom number,
$N(t) = N(0)e^{-\gamma t}$, with $1/\gamma=3.0(1)$\,s.
The effect of such one-body loss has a simple expression for two-mode squeezed states \cite{Li2008}, $\xi_N^2(t)-1 = (\xi_N^2(0)-1)e^{-\gamma t}$. This result constitutes a fundamental lower bound for the time-dependent spin squeezing in a real system.
To go further,
the dynamics of the full internal and external quantum state need to be taken into account; in particular, the measurement-induced light shift and spin-exchange interactions. 
First, the cavity measurement leaves behind a phase shift $\phi\propto S_z$ depending on the measured $S_z$ \cite{Schleier-Smith2010a}. This is the driving force for cavity feedback squeezing \cite{Leroux2010} but is usually neglected in measurement-based squeezing \cite{Schleier-Smith2010a,Hosten2016}. However, for the long interaction times in our experiment, it conspires with spin-exchange interaction to give rise to a mechanism explaining the observed amplification. Indeed, for a given atom $i$, this phase shift also depends on its coupling $\Omega_i$, so that we have $\phi_i\propto\Omega_i S_z$. As $\Omega_i$ in turn depends on the transverse motional energy of the atom, this corresponds to a spin-orbit coupling where atoms with small oscillation amplitude experience above-average phase shift. This correlation persists until collisions redistribute motional energy, i.e., for a time on the order of 3\,s in our experiment (see Table~\ref{tab:1}).
Second, spin-exchange interaction rotates individual spins about the axis of total spin at rate $\omex$, thus converting the phase-shift deviation $\delta \phi_i$ of an atom from the ensemble mean into population difference $\delta s_{z,i}$ \cite{Deutsch2010,Kleine2011,Solaro2016}.
(Note, however, that the squeezed axis remains unchanged---i.e., along $z$---because the interaction can be described by a Hamiltonian $\propto\hat{\bf{S}}\cdot\hat{\bf{S}}$ which is a constant of motion that commutes with $\hat{S}_z$ \cite{Martin2013}.)
While this interaction conserves total spin, it does convert the initial correlation $\delta \phi_i\propto\Omega_i$ into a correlation $\delta s_{z,i}\propto\Omega_i$, which will affect the result of a subsequent cavity measurement. In the case of our measurement scheme, the spin-rotation direction is such that for $t<\pi/\omex$, atoms with above-average coupling ($\Omega_i>\Omega_e$) acquire $\delta s_{z,i}>0$ if $S_z>0$ and vice versa:
strongly coupled atoms acquire an increased population difference, $\avg{s_{z,i}}|_{\Omega_i>\Omega_e} = AS_z/N$ with $A>1$. As these atoms make a dominant contribution to the cavity measurement, a second cavity measurement is amplified with respect to $M_1$.

\begin{figure}[tb]
\includegraphics[width=0.4\textwidth]{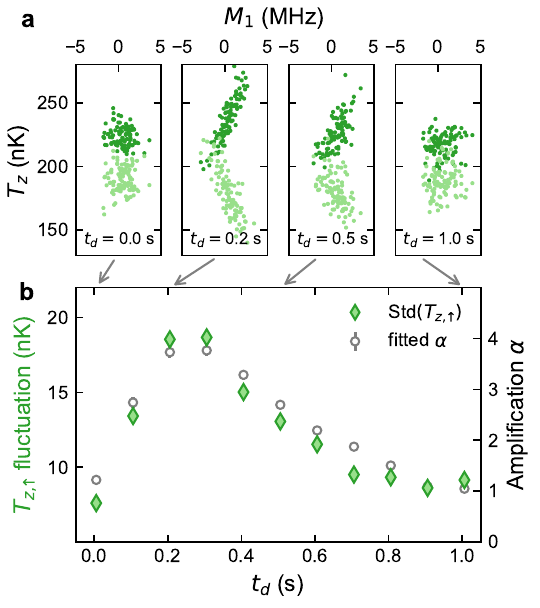}
\caption{\label{fig:temp_corr} {The spin-orbit correlation from temperature measurements.}
\textbf{a}, The temperatures along $\hat{z}$ of states $\kdn$ (light green dots) and $\kup$ (dark green dots) versus $M_1$ from the same data set (the offset between the initial temperatures is due to the state-dependent imaging procedure). There is a clear correlation between $T_{z,\uparrow}$ and $M_1$ (anticorrelation for $\kdn$) for $t_d=0.2\,$s and 0.5\,s, where $\alpha>1$.  
\textbf{b}, The time evolution of the temperature fluctuations (standard deviation) of $T_{z,\uparrow}$. The evolution closely resembles that of the amplification factor $\alpha$ (open circles, right axis), further corroborating the amplification mechanism described in the text.}
\end{figure}

The time-of-flight imaging yields state-resolved temperatures $\Tup$ and $\Tdn$ for every shot, providing an experimental test for this mechanism. If $s_{z,i}$ is correlated with $\Omega_i$ and hence with the motional energy as outlined above, then the temperatures of the two spin components should be correlated with $M_1$ and the correlation should have opposite signs for the two states (Appendix \ref{app:model}). Indeed, when $\Tup$
and $\Tdn$ are plotted against $M_1$ [Fig.~\ref{fig:temp_corr}(a)], a correlation is clearly visible for times $t_d$ where $\alpha>1$ and has the expected sign: the higher the measured $S_z$ in $M_1$, the lower is $\Tdn$ (taking into account the base change ($\pi$ pulse) in $M_2$) and the higher is $\Tup$. The amount of temperature change also depends on
$\alpha$ in the expected manner, as can be seen by plotting the shot-to-shot temperature fluctuations (standard deviation $\Delta T_z$) as a function of $t_d$ [Fig.~\ref{fig:temp_corr}(b), green diamonds]. For short times where $\alpha\sim 1$, fluctuations are very low, limited by measurement noise. As $\alpha$ increases, they rise up to approximately $18\,$nK, and their time evolution closely follows that of $\alpha$ for our measurement time.

A semiclassical Monte Carlo simulation where atoms move on classical trajectories and evolve under mean-field spin exchange equation (Appendix \ref{app:nsim}) reproduces the time evolution of the amplification factor quite well, as shown in Fig.~\ref{fig:corr_t}(b). It also reproduces the trend of the spin-dependent temperature evolution. The simulation includes a damping rate that is estimated from the decay rate of center-of-mass oscillations mostly induced by the cavity-locking light (Appendix~\ref{app:nsim}). The quantitative agreement is satisfactory despite the simplicity of the model and the fact that the simulations do not take quantum correlations into account.

The observed temperature correlation and the simulation results thus provide strong evidence for the amplification effect resulting from the inhomogeneous measurement-induced phase shift combined with spin exchange interaction, acting on the squeezed state on the long time scales explored here for the first time. This is in contrast to the noninteracting case usually considered, where differences in atom-cavity coupling merely reduce the effective atom number \cite{Hu2015} and a small dephasing does not affect the measurement as it remains confined to an axis that is not observed. Also note the difference with respect to ``quantum phase magnification'' effects \cite{Davis2016,Hosten2016s}: while in these effects, interactions modify the spin state itself, here it is the correlations between the spin and motional degrees of freedom that are modified and lead to the observed amplification.

\begin{figure}[tb]
\includegraphics[width=0.4\textwidth]{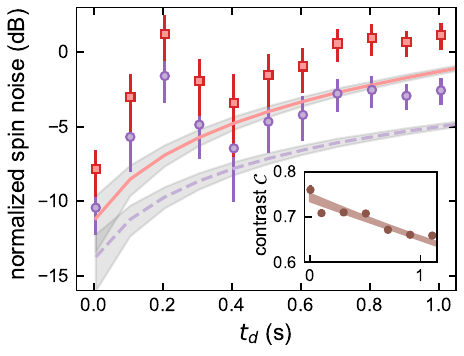}
 \caption{\label{fig:sqz_t} {Inferred squeezing in the spin degree of freedom.} The purple circles show the inferred upper limit of number squeezing in the spin degree of freedom (Eq.~\ref{eq:noise_sz_fin}). The evolution is consistent with the theoretical lower bound given by atom loss (dashed curve). The red squares show the resulting metrological squeezing, obtained as usual by dividing the number squeezing by the measured Ramsey contrast shown in the inset.
 For comparison, the solid curve is the theoretical limit normalized to the fit of the experimental Ramsey contrast. The inset shows the Ramsey contrast data (circles) as a function of $t_d$ and their exponential fit (shaded curve) which yields $\tau=7.7(6)\,$s.
 }
\end{figure}

Based on this understanding, an upper limit of the spin noise at time $t$ in presence of atom loss (rate $\gamma$) and amplification [factor $\alpha(t)$]  can be deduced from the measurements (see Appendix~\ref{app:noise_model}):
\begin{align}
     \!\!\!\!\Delta^2 S_z(t)|_{M_1} &\!= \Omega_e^{-2}[ \Var\left(M_2 - \alpha M_1\right) \nonumber \\
     &-(\alpha^2-e^{-2\gamma t})\Delta^2 M_1^\rm{psn}-\Delta^2 M_2^\rm{psn}]\,,
\label{eq:noise_sz_fin}
\end{align}
where we drop the time dependence of $\alpha$ and $M_2$ for simplicity and the only noise we subtract is the PSN, $\Delta^2 M_l^\rm{psn}$. We find that the contribution from the noise of $M_1$ (second term in the square brackets) is also affected by the amplification mechanism ($\alpha^2$) and the exponential comes from the decay of the total spin.
This upper limit (purple circles in Fig.~\ref{fig:sqz_t}) approaches the lower limit due to losses (dashed line in Fig.~\ref{fig:sqz_t}) to within 3\,dB  except in the time interval where $\alpha$ is significantly larger than 1.
The larger difference at those times may come from fluctuation in the amplification dynamics themselves. 
We also plot the metrological squeezing factor by combining these data with the independently measured coherence as a function of time (Fig.~\ref{fig:sqz_t}, inset).
The inferred metrological squeezing remains below 0\,dB up to 600\,ms. This is about 2 orders of magnitude longer than in previous squeezing experiments with cold alkali atoms, and compatible with the typical interrogation time for primary-standard atomic clocks.
The next steps in our experiment will be to devise more symmetric measurement pulse schemes as discussed in Sec.~\ref{sec:concl}, and to implement the real-time feedback that allows measurement-based squeezing to be integrated into the full Ramsey-measurement cycle.

\section{Discussion and conclusion}\label{sec:concl}

The observed amplification mechanism due to a spin-orbit correlation reflects the rich dynamics in the squeezed clock setting and shows that these dynamics need to be taken into account in metrological applications of spin squeezing. Two major factors govern the amplification effect---the exchange rate $\omex$ and the inhomogeneity in the coupling $\Omega_i$. The same ingredients are present in other cavity-based spin-squeezing systems and typically are on the same order of magnitude. They are also present in other interacting spin systems, where they also lead to nontrivial many-body physics \cite{Smale2019,Norcia2018b,Davis2019}. 
Our results thus show that the inhomogeneous measurement back action on phase is a crucial factor for metrological sensors with alkaline atoms.
 While the amplification effect in its current form ($\alpha>1$) is remarkable from a conceptual point of view, the fact that it also amplifies the fluctuations in the first measurement may limit its usefulness for quantum enhanced phase measurement. However, our model also reveals how the amplification effect can be controlled. By engineering the measurement-induced phase shift, it is possible to change the sign of the time dynamics of $\alpha(t)$, or even reduce $\alpha(t)$ to zero at a particular time, in which case the model suggests that the final phase measurement could be done as in deterministic squeezing, without being affected by the result of the preparation measurement. Alternatively, it should be possible to devise more symmetric measurement schemes where the mean light shift is always zero, even when $S_z\neq0$.
Most importantly, however, the experimental results shown here demonstrate that the spin-squeezing lifetimes required for metrology-grade clocks and sensors can be experimentally achieved in spite of real-life effects such as inhomogenous coupling and spin interactions.

\section*{Acknowledgments}
\label{sec:ack}
We thank Jean-No\"el Fuchs, Fr\'ed\'eric Pi\'echon and Franck Lalo\"e for fruitful discussions, and also thank J.-N. F. for assistance with the numerical simulation. We acknowledge early contributions to the experiment from Ralf Kohlhaas and Th\'eo Laudat and are indebted to Tobias Gross of Laseroptik GmbH for advice on the fiber mirror coatings.
This work is supported by the European Research Council (ERC) (Advanced Grant 671133 EQUEMI), by the D\'el\'egation G\'en\'erale  de l'Armement (DGA) via the ANR ASTRID program (Contract No. ANR-14-ASTR-0010) and by R\'egion Ile-de-France (DIM SIRTEQ).

\appendix

\section{\label{app:exp}Experimental Details}

\subsection{\label{app:timescales}Time scales}
Table~\ref{tab:1} summarizes the relevant time scales involved in the experiment.
\begin{table}[h]
\begin{tabular}{ccc}
    \hline
    Transverse trap frequency & $\omega_y$, $\omega_z$ & $\sim 120$\,Hz \\
    Longitudinal trap frequency & $\omega_x$ & $\sim 7.5$\,Hz \\
    Spin-exchange rate & $\omex/2\pi=2\hbar|\aupdn|\bar{n}/m$ & $\sim 1\,$Hz \\
    Lateral collision rate & $\gamma_c=(32\sqrt{\pi}/3)\aupdn^2\bar{n}v_T$ & $\sim 0.3\,$Hz \\
    Background loss rate & $\gamma$ & $\sim0.33\,$Hz \\
    Phase-decoherence time & & $\sim0.05\,$Hz \\
    \hline
   \end{tabular}
\caption{A summary of the experimental time scales, in which $\bar{n}\sim 1.6\times10^{11}\,\rm{cm}^{-3}$ is the average atomic density, $\aupdn\approx 98.09a_0$, with $a_0=0.0529\,$nm, is the relevant scattering length, $m$ is the atomic mass, $v_T$ is the thermal velocity $v_T\approx\sqrt{k_BT/m}$ and $k_B$ is the Boltzmann constant. }
\label{tab:1}
\end{table}

\subsection{\label{app:setup}Setup parameters}
\begin{figure}[tb]
\includegraphics[width=0.4\textwidth]{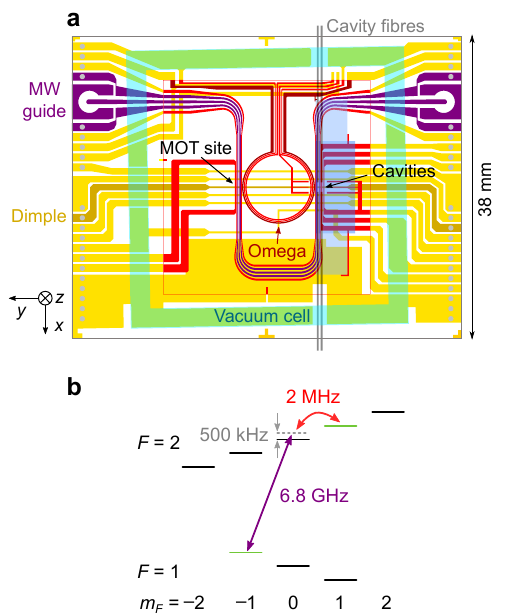}
\caption{\label{fig:chip}\textbf{a}, The layout of the atom chip, which has a two-layer structure and incorporates two fiber Fabry-Perot (FFP) cavities of different finesse. Only the low-finesse cavity (located left in the image) is used in the experiments described here. The red (yellow) wires are on the top (bottom) chip. The central ``Omega'' wire is used to transfer atoms from the MOT site to the cavity. The green shading indicates the cross section of the vacuum cell. \textbf{b}, The Zeeman levels of the $^{87}$Rb ground state. The clock states are marked in green. The clock transition is excited with two-photon pulses as indicated.}
\end{figure}

The layout of the atom chip and details of
the two-photon clock transition are shown in Fig.~\ref{fig:chip}. The microwave (MW) photon is detuned $454\,$kHz
above the $\kdn\rightarrow\ket{F=2,m_F=0}$ transition and is delivered by an on-chip coplanar waveguide. The radio-frequency (rf) photon is delivered from another chip wire.
After magneto-optical trapping (MOT), optical molasses and optical pumping
to $\kdn$, atoms are magnetically trapped at the MOT site and magnetically transported to the cavity using the ``omega wire'' (Fig.~\ref{fig:chip}), where the trap is compressed and forced rf evaporative cooling is applied. Finally, the trap is decompressed to its final parameters (``interrrogation trap'') and positioned exactly inside the optical cavity mode. Due to the low density, the final state is not completely thermalized and has a slight temperature difference between the longitudinal and transverse axes, as quoted in the main text. The complete loading and preparation phase takes 3\,s.
In the interrogation trap, the magnetic field at the bottom of the trap points along $\hat{x}$ and has a value $B_x=B_m-35\,$mG, where $B_m= 3.229\,$G is the ``magic'' field for which the linear differential Zeeman shift between the clock states vanishes \cite{Harber2002} and the 35-mG offset maximizes the coherence time \cite{Deutsch2010}.

The state-resolved imaging starts with a MW pulse that adiabatically transfers atoms from $\kdn$ to $\ket{F=2,m_F=0}$, where they are no longer trapped and start to fall. The trap is turned off several milliseconds later to release atoms in $\kup$, such that the two clouds are well separated and are imaged in a single picture. However, the adiabatic transfer also perturbs the trap so that the temperature estimation is slightly biased.

The optical cavity is symmetric with a finesse $\mathcal{F}=2.7(1)\times10^3$ for the 780-nm mode. This gives a maximum single-atom cooperativity $C_0=24\mathcal{F}/\pi k_{780}^2w_0^2\approx1.9$, where $k_{780}$ is the wave vector of the probe laser. Taking into account the inhomogeneity for our trapped cloud with $T\sim 200\,$nK, the effective cooperativity is $C_\rm{eff}\approx 0.42$.
The cavity is simultaneously resonant for a stabilization wavelength at 1560\,nm. The stabilization laser is constantly on during the experiment but its intracavity intensity is sufficiently weak to prevent trapping of the atoms (trap depth $\lesssim20\,$nK).

\subsection{\label{app:calib}Calibrations}

The imaging system is calibrated using the known $\sqrt{N}$ scaling of the projection noise of a coherent state, similar to Ref.~\cite{Riedel2010}.
To measure $\Omega_e$, we prepare CSSs with different $\avg{S_z}$ by applying a weak MW+rf pulse of variable length. Cavity transmission spectra are obtained by scanning a weak probe laser over 20 cavity line widths in 50\,ms. We obtain the prepared $\avg{S_z}$ from the imaging data.
A linear fit of the cavity frequency versus the prepared $\avg{S_z}$ yields $\Omega_e$. Our preparation procedure leads to a small dependence between the temperature and the prepared atom number. Therefore, the measured $\Omega_e$ depends slightly on $N$ ($ 1.5$\% deviation for 10\% change in $N$).

We calibrate the phase shift induced by the cavity probe using a Ramsey sequence (with the probe pulse occurring during the Ramsey time). We obtain the ensemble-average phase shift per detected photon $\bar{\phi}_d = 4.16(2)\times10^{-4}\,\pi\,$rad. Ideally, for a given atom $i$, the phase shift is given by $\phi_{i} = \frac{\Omega_i}{\kappa_t} n_t$, where $n_t$ is the transmitted photon number and $\kappa_t = \mathcal{T}c/(2L) \lesssim \kappa/2$ is the transmission rate, in which $\mathcal{T}=1000$ parts per million is the designed mirror transmission and $c$ the speed of light. This allows us to estimate the overall photon detection efficiency $\eta$ by comparing $\bar{\phi}_d$ with the expected phase shift per \emph{transmitted} photon ($\avg{\phi_i}/n_t = \Omega_e/\kappa_t$).

\subsection{\label{app:composite}Composite cavity measurement}

We define the composite cavity measurement $M_l=(\delta\omega_{l+}-\delta\omega_{l-})/2$ ($l=1,2$). In order to account for the $\pi$ pulses that flip $S_z$, we define $\delta \omega_\pm$ accordingly, such that $\delta \omega_+$ ($\delta\omega_-$) refers to the second (first) probe for $M_1$ but refers to the first (second) probe for $M_2$ [see Fig.~\ref{fig:scheme}(d)]. Consequently, $S_z$ refers to the state after $M_1$.

We obtain the cavity shifts $\delta\omega_\pm$ from the transmitted photon number, taking into account the Lorentzian line shape.
At the end of each experimental cycle (after atoms are imaged), we apply two additional cavity-probe pulses with $\pm\kappa/2$ detuning, to calibrate possible long-term drift of the cavity frequency and the probe intensity.

Experimentally, we employ a SCROFULOUS composite $\pi$ pulse \cite{Cummins2003}, with each of the constituent pulses tuned to a duration of the transverse trap period $2\pi/\omega_z$.
This helps to reduce the pulse error due to amplitude inhomogeneity and fluctuation.

\section{\label{app:inhomo}Inhomogeneous coupling}

The atom-field coupling in the cavity is a function of the atomic trajectory $\bm{r}(t)=\lbrace x,y,z\rbrace$ and is determined by the cavity geometry:
\begin{equation}
    \Omega(\bm{r}) = \Omega_0\cos^2\left(k_{780}x\right)\left(\frac{w_0}{w}\right)\exp\left[-2\frac{y^2+z^2}{w^2}\right]\,,
    \label{eq:cav_coupling}
\end{equation}
where $w=w_0\sqrt{1 + x^2/L_R^2}$, in which $L_R=k_{780} w_0^2/2\approx750\,\mu$m the Rayleigh length.
The maximum shift $\Omega_0$ can be obtained from the experimentally measured $\Omega_e$ and agrees with the value obtained from a cavity quantum electrodynamics calculation.
The time integral of $\Omega(\bm{r})$ over the pulse duration $\tau_p$ yields the effective coupling $\Omega_i = \frac{1}{\tau_p}\int_0^{\tau_p} dt \Omega(\bm{r}_i)$ for atom $i$ used in the main text.
Assuming harmonic oscillation, the position dependence in the transverse directions reduces to a function of the motional energies $E_{y,i}$ and $E_{z,i}$:
\begin{align}
    \Omega_i \approx \Omega_0 \left( 1 - \frac{x_i^2}{L_R^2}\right) e^{-\left(\frac{E_{y,i}}{\varepsilon_y}+\frac{E_{z,i}}{\varepsilon_z}\right)}I_0\left(\frac{E_{y,i}}{\varepsilon_y}\right)I_0\left(\frac{E_{z,i}}{\varepsilon_z}\right)\,,
\label{eq:inhomo}
\end{align}
where $\varepsilon_y \equiv \frac{m\omega_y^2w_0^2}{2}$ and $\varepsilon_z \equiv \frac{m\omega_z^2w_0^2}{2}$.
In the experiment, the averaging is not perfect along $\hat{y}$ since $\omega_y $ and $\omega_z$ are not precisely equal. $I_0(\cdot)$ is the modified Bessel function of the first kind. Note that we assume the standing wave in $\hat{x}$ can be averaged out and that the position dependence on $x$ is weak as the cloud size $\ll L_R$.
As a result, as in most real systems, atoms contribute differently to the quantum fluctuations of $\delta\omega=\Omega_e S_z$.
Nevertheless, the system can be described as a uniformly coupled one with a slightly reduced effective atom number $N_\rm{eff} = \frac{(\sum_i^N\Omega_i)^2}{\sum_i^N\Omega_i^2}$ and coupling $\Omega_\rm{eff}=\frac{\sum_i^N\Omega_i^2}{\sum_i^N\Omega_i}$, as long as the couplings do not change over time \cite{Schleier-Smith2010a,Hu2015}. Note that as $N_\rm{eff}\Omega_\rm{eff} = N\Omega_e$ and $\xi^2=4(\Delta S_z)^2\vert_{M_1}/(N\mathcal{C}^2)\propto 1/N\Omega_e^2$ (cf. Eq.~\ref{eq:cond_sz}), the squeezing will appear higher if $N_\rm{eff}$ and $\Omega_\rm{eff}$ are used. For our system, $N_\rm{eff}\approx 0.90N$ and $\Omega_\rm{eff}\approx 1.11\Omega_e$, so that the effect on $\xi^2$ is within 10\%.
We use $N$ (measured by imaging) and $\Omega_e$ to obtain a conservative estimate of the squeezing.

\section{Data analysis}
\subsection{\label{app:coherence}Coherence measurements}
To determine the atomic coherence after a composite measurement [Fig.~\ref{fig:sqz}(b) and Fig.~\ref{fig:sqz_t}, inset], we apply a second $\pi/2$ pulse after $M_1$, forming a Ramsey sequence with $M_1$ occurring during the Ramsey time. By varying the phase of the second $\pi/2$ pulse, we obtain Ramsey fringes ($S_z$ versus phase). However, $M_1$ induces an average phase shift depending on the measured $S_z$ value, which fluctuates from shot to shot due to quantum projection noise. We correct this phase shift in the data analysis using the calibrated phase shift per detected photon (Appendix~\ref{app:calib}) and the number of detected photons in $M_1$ in each shot. We can then obtain the contrast with a sinusoidal fit to the Ramsey fringes.

We fit the contrast decay as a function of average detected photons to $\mathcal{C}=\exp[-\avg{n_1}/\gamma_1 -\left(\avg{n_1}/\gamma_2\right)^2]$ [Fig.~\ref{fig:sqz}(b)], yielding $\gamma_1 = 3(1)\times10^5\gg\gamma_2 = 1.88(7)\times10^4$. The second term dominates, which can be understood as follows. The imperfection in the spin-echo compensation leads to a Gaussian distribution of the atomic phase, the width of which depends linearly on the measurement strength ($\avg{n_1}$). This gives the dominant scaling $\mathcal{C}\propto\exp[-\avg{n_1}^2]$ The imperfection may arise from the spin dynamics occurring during the spin-echo sequence, the residual inhomogeneous coupling along the cavity axis, and the infidelity of the $\pi$ pulse.

\subsection{\label{app:sqz} Conditional noise}
In Eq.~1, the variance of $M_2$ conditioned on $M_1$, $\rm{Var}(M_2-\zeta M_1)$ is minimized by $\zeta = \rm{Cov}(M_1,M_2)/\rm{Var}(M_2)$, equivalent to the slope of a simple linear regression between $M_2$ and $M_1$ ($M_1$ being the independent variable).
It is worth noting that even if $M_1$ and $M_2$ are identical measurements (same atom-cavity coupling), $\zeta$ deviates from 1 when there is error in $M_1$ (such as PSN). Nevertheless, Eq.~1 remains the conventional way to evaluate the conditional spin noise, except that $\zeta$ is slightly biased for determining the true relation between $M_2$ and $M_1$.
For consistency and simplicity, we use the same linear regression (with $M_1$ as the independent variable) to obtain the amplification factor $\alpha$ [Fig.~3(a)], which minimizes the variance in Eq.~\ref{eq:noise_sz_fin}. The bias is almost negligible.

In Fig.~\ref{fig:corr_t}(c), we show a conditional variance $\rm{Var}(M_1-M_2/\alpha)$, seemingly different from the conditional variance in Eq.~\ref{eq:cond_sz}. It is, in fact, the same formula given that we aim to find what one learns about $M_1$ from $M_2$; in other words, a retrodiction of $M_1$ knowing $M_2$ or a conditional variance of $M_1$ given $M_2$. As we mentioned, this is a simple way to quantify the correlation without knowledge of how the amplification effect might influence the spin. 

In Sec.~\ref{sec:ampli}, with further knowledge about the amplification effect, we infer the spin squeezing using Eq.~\ref{eq:cond_sz} properly, leading to Eq.~\ref{eq:noise_sz_fin}. It is based on the conditional variance of $M_2$ given $M_1$ as a phase measurement in a clock application would be.

\subsection{\label{app:tomo}Spin tomography}
$\Delta S_\theta^2$ is also estimated in a conditional way similar to Eq.~\ref{eq:cond_sz} (as in \cite{Cox2016}):
\begin{equation}
    (\Delta S_z)^2_\theta \leq \left[\rm{Var}(M_1\cos\theta  - M_2) - (\Delta M_2^\rm{psn})^2\right]/\Omega_e^2\,.
\end{equation}
The data shown in Fig.~\ref{fig:sqz}(c) are after a postselection of the measured $S_z$ (close to 0), because with our composite measurement scheme, the shot-to-shot phase fluctuation is dominated by the quantum fluctuation of $S_z$ (see Eq.~\ref{eq:phaseshift} below). In principle, this phase fluctuation can be suppressed by an active feedback on the phase based on the cavity measurement result, up to the ultimate PSN. Postselection simulates the optimal situation with active feedback, while the discrepancy between the data and the prediction (pink curve) comes from the fact that the postselection is not stringent due to the limited number of samples.
With an optimal phase feedback, PSN induces at least 6.1\,dB excess antisqueezing, which needs to be taken into account in real clock applications \cite{Braverman2018}.
It is worth noting that cavity feedback squeezing \cite{Leroux2010}, which can enable near unitary squeezing \cite{Braverman2019}, can also be implemented in our system.

\subsection{\label{app:noise_model} Spin noise model under the amplification effect}

From the qualitative understanding of the amplification mechanism outlined in the main text and supported by the temperature correlation with $M_1$ (Fig.~\ref{fig:temp_corr}), we can formulate a simple phenomenological model of the time evolution of the cavity measurement. 

We start by formulating the time evolution of $S_z$ given a measurement $M_1$ as: 
\begin{equation}
    S_z(t)|_{M_1} = e^{-\gamma t}S_z(0)|_{M_1} + \delta S_z(t)\,,
\end{equation}
where $S_z(0)|_{M_1}$ follows the conditional probability distribution of $S_z$ after $M_1$, i.e., approximately a normal distribution centered around $M_1/\Omega_e$ with a variance given by $\rm{Var}(\delta M_1^\rm{n})/\Omega_e^2$. (We use $\delta M_l^\rm{n}$ to represent the noise of measurement $l=1,2$ which is much lower than atomic projection noise, and the lower bound of which is given by photon shot noise $\rm{Var}(\delta M_l^\rm{n})\geq\Delta^2 M_l^\textrm{psn}$.) $\delta S_z(t)$ represents all spin noise occurring after $M_1$ (such as the loss-induced noise \cite{Li2008}); $e^{-\gamma t}$ accounts for the reduction of the spin $\avg{S_z}$ due to one-body loss. The spin variance is thus given by
\begin{align}
    \Delta^2 S_z(t)|_{M_1} =  e^{-2\gamma t}\rm{Var}(\delta M_1^\rm{n})/\Omega_e^2 + \Var(\delta S_z(t))\,.
    \label{eq:sz_t}
\end{align}
On the other hand, we expect $M_2$ to follow a statistical distribution given by 
\begin{equation}
 M_2(t) = \Omega_e\left[ \alpha(t)S_z(0)|_{M_1}+\delta S_z(t)\right]+\delta M_2^\rm{n}\,,
\label{eq:noise}
\end{equation}
where $\alpha=\alpha'e^{-\gamma t}$ includes both the pure amplification effect $\alpha'$ and the reduction of the spin $\avg{S_z}$ due to one-body loss. Note that the amplification mechanism acts on the phase correlation imprinted in $M_1$ but does not modify the cavity coupling of the spins.

To infer $\Delta^2 S_z(t)|_{M_1}$ from our cavity measurements $M_1$ and $M_2$, we use the minimum conditional variance as in Eq.~\ref{eq:cond_sz} ($\zeta=\alpha$ minimizes the variance). Assuming the three contributions in Eq.~\ref{eq:noise} to be statistically independent, we have
\begin{multline}
  \Var\left(M_2(t)- \alpha(t) M_1\right) =\\
  \alpha(t)^2 \rm{Var}(\delta M_1^\rm{n})+ \Omega_e^2\Var(\delta S_z(t)) + \rm{Var}(\delta M_2^\rm{n})\,.
  \label{eq:corr_t}
\end{multline}
We find that this variance does contain information about $\Var[\delta S_z(t)]$ but it is affected by the noise of $M_1$ amplified by $\alpha(t)$ (first term on the right-hand side). Nevertheless, knowing $\alpha(t)$ from our data allows us to infer the actual spin noise $\Delta^2 S_z(t)|_{M_1}$ (Eq.~\ref{eq:sz_t}, i.e., the correlation only in the spin degree of freedom rather than the combined spin-orbit observable seen by $M_2$) by comparing it with Eq.~\ref{eq:corr_t}:
\begin{align}
     \Delta^2 S_z(t)|_{M_1} &\!= \Omega_e^{-2}[ \Var\left(M_2(t) - \alpha(t) M_1\right) \nonumber \\
     &-(\alpha(t)^2-e^{-2\gamma t})\rm{Var}(\delta M_1^\rm{n})-\rm{Var}(\delta M_2^\rm{n})]\,.
\end{align}
Taking a conservative limit by assuming the minimum PSN from the cavity measurements, we obtain Eq.~\ref{eq:noise_sz_fin}.

With this model, we can also calculate the other conditional variance $\rm{Var}(M_1-M_2/\alpha)$ that we use to evaluate the correlation [Fig.~\ref{fig:corr_t}(c)] without knowing the model. Similarly we have
\begin{multline}
  \Var\left(M_1- M_2/\alpha\right) =\\
   \rm{Var}(\delta M_1^\rm{n}) + \rm{Var}(\delta M_2^\rm{n})/\alpha^2 + \Omega_e^2\Var(\delta S_z(t))/\alpha^2\\
   \geq\rm{Var}(\delta M_1^\rm{n}) + \frac{\rm{Var}(\delta M_2^\rm{n})}{\alpha^2} + \frac{\Omega_e^2 N(t)}{4\alpha^2}(1-e^{-\gamma t})\,,
  \label{eq:retro_theory}
\end{multline}
where in the last line we use the minimum spin noise $\delta S_z(t)$ caused by one-body loss, derived from $\xi_N^2(t)-1 = (\xi_N^2(0)-1)e^{-\gamma t}$ \cite{Li2008}. Eq.~\ref{eq:retro_theory} is plotted in Fig.~\ref{fig:corr_t}(c), after normalizing to the SQL at $t=0$, $\Omega_e^2 N(0)/4$.
The agreement with data therefore supports our model of the spin noise. 

\section{Amplification mechanism}
\subsection{\label{app:model} Semiclassical model}

Here, we formulate a simple semiclassical model that reproduces the amplification effect. We make the following assumptions. (1) $\Omega_i$ is only determined by $E_{y,i}$ and $E_{z,i}$, which are conserved during the experiment (Eq.~\ref{eq:inhomo}). The ensemble coupling $\Omega_e = \frac{1}{N}\sum_i^N \Omega_i$ is then a constant. (2) The spin rotation is modeled as a simple rotation of each spin around the ensemble average with the same rate $\mathcal{C}\omega_\rm{ex}$, determined by the atomic coherence. We ignore other sources of dephasing, such as dephasing from the trapping potential. (3) We also assume a perfect $\pi$ pulse on the clock transition for the spin echo and no spin dynamics during the composite measurement.

The phase shift induced by $M_1$ is obtained from the transmitted photon numbers $n_{1\pm}$ in the two probe pulses.
With a linear approximation of the cavity transmission (probe detuning $\kappa/2$), $n_{1\pm} \approx n_p ( 1+ 2\delta\omega_{1\pm}/\kappa)$, where $n_p$ is the average transmitted photon number per probe pulse (an experimental parameter), $n_p = \avg{n_1}/(2\eta)$ and $\avg{n_1}$ is the average \emph{detected} photon number in $M_1$ used in the main text. According to our sign convention (Appendix~\ref{app:composite}), the first probe gives $\phi_{i-} = \frac{\Omega_i}{\kappa_t}n_p\left(1 -\frac{2\Omega_e}{\kappa}S_z\right)$ (note the sign of $S_z$, which acquires another minus sign after the spin-echo pulse). The second probe gives $\phi_{i+} = \frac{\Omega_i}{\kappa_t}n_p\left(1 +\frac{2\Omega_e}{\kappa}S_z\right)$ and the total phase shift in $M_1$ reads
\begin{equation}
\phi_i = \phi_{i+} - \phi_{i-} =
\frac{4\Omega_e n_p}{\kappa_t\kappa}\Omega_i S_z
\label{eq:phaseshift}
\end{equation}
The phase deviation from the mean phase
$\bar{\phi} = \arctan\left(\sum_{i}\sin(\phi_i)/ \sum_{i}\cos(\phi_i)\right)\approx\frac{1}{N}\sum_i\phi_i$ is then
\begin{equation}
\delta_i = \phi_i - \bar{\phi} = \chi S_z(\Omega_i - \Omega_e)\,,
\end{equation}
where $\chi =\frac{4\Omega_e n_p}{\kappa_t\kappa}\approx \frac{4\Omega_e\avg{n_1}}{\eta\kappa^2}$. This phase distribution would not be measurable by the cavity until that spin-exchange collisions rotate inividual spins about the total spin [known as the identical spin-rotation effect (ISRE) \cite{Piechon2009,Deutsch2010}].
The effect of the ISRE is then to rotate individual spins about their sum. The rotation rate is determined by $\mathcal{C}\omex$.
While the total $S_z$ is conserved, the $s_z$ values of individual atoms evolve as
\begin{equation}
  s_{z,i}(t)=s_z^0+\frac{\delta_i}{2}\sin\mathcal{C}\omex t\,.
  \label{eq:s_zi}
\end{equation}
The initial value is close to $s_z^0=S_z/N$ for all atoms due to the QND measurement and the plus sign is determined by the relevant scattering lengths in $^{87}$Rb \cite{Fuchs2002}. We then obtain cavity shift of $M_2$:
 \begin{equation}
\delta\omega(t) = \sum_i^N\Omega_i s_{z,i}(t) = \Omega_e S_z \left(1+a_m \sin\mathcal{C}\omex t \right)\,,
\label{eq:cav_amp}
\end{equation}
where $a_m=\chi N (\Delta\Omega)^2/2\Omega_e$ and $(\Delta\Omega)^2=\frac{1}{N}\sum_i^N(\Omega_{i}-\Omega_e)^2$ is the variance of the coupling. We find that
\begin{equation}
  \alpha(t)=1+a_m \sin\mathcal{C}\omex t  
  \label{eq:alpha_model}
\end{equation}
is the time-dependent amplification factor.
We thus expect an amplification that depends on the atom number, the probe photon number,
and the coupling inhomogeneity and that increases for $t\lesssim \pi/(2\mathcal{C}\omex)$. While this simplified model predicts an oscillation of $\alpha(t)$, we expect it to damp out for times approaching the lateral collision time scale, as these collisions destroy the correlation between motional and internal state.

The model also predicts that a correlation should arise between the spin state and motional energy. For example, when $M_1$ yields $S_z>0$, the ISRE converts the phase shift of colder atoms
into an increased probability of being in $\kup$ and that of hotter atoms into an increased probability of $\kdn$, for times $t<\pi/(\mathcal{C}\omex)$.
More quantitatively, we consider the motional energy $E_{t,i}$ of atom $i$ in the transverse directions ($t=y,z$). $\Omega_i$ is a monotonically decreasing function of $E_{t,i}$ (see Eq.~\ref{eq:inhomo}) and here we approximate it by $\Omega_i - \Omega_e\approx -\varepsilon(E_{t,i}-\bar{E_t})$, where $\bar{E_t} = \frac{1}{N}\sum_i^N E_{t,i} = k_BT$ and $\varepsilon$ is a positive constant. It follows that $\rm{Var}(\Omega) \approx \varepsilon^2 \rm{Var}(E_t) = \varepsilon^2(k_BT)^2$, so $\varepsilon=\Delta\Omega/k_BT$.
Overall, the average energy of $\kup$ can be written as $E_{t,\uparrow}\approx\frac{1}{N_{\uparrow}}\sum_i^N P_{\uparrow,i}E_{t,i}$, where $P_{\uparrow,i}=\frac{1}{2}- s_{z,i}$ and $N_{\uparrow} = \sum_i^N P_{\uparrow,i} = N/2-S_z$ (and similarly for $\kdn$, with $P_{\downarrow,i}=\frac{1}{2}+ s_{z,i}$). Note the replacement $s_z\rightarrow -s_z$ due the final base change ($\pi$ pulse) in $M_2$. The ISRE furthermore correlates $s_{z,i}$ with $E_{t,i}$ through $\Omega_i$.
Using Eq.~\ref{eq:s_zi}, after an evolution time $t$,
\begin{align}
    E_{t,\uparrow}\approx&~ \frac{2}{N - 2S_z}\sum_i^N\left(\frac{1}{2}-s_{z,i}\right)E_{t,i} \nonumber\\
    \approx&~\bar{E_t} +  \frac{\chi S_z \sin\mathcal{C}\omex t}{N-2S_z}\sum_i^N \varepsilon(E_{t,i} - \bar{E_t})E_{t,i} \nonumber\\
    \approx&~ k_BT + \chi\varepsilon(k_BT)^2S_z\sin\mathcal{C}\omex t
\end{align}
where we use $N\gg S_z$; and $\rm{Var}(E_t) = (k_BT)^2$ for thermal distribution.
The experimentally measured transverse temperature directly links to the average energy as $T_{t,\uparrow(\downarrow)} \approx E_{t,\uparrow(\downarrow)}/k_B$. This leads to
\begin{equation}
    T_{t,\uparrow} \approx T\left(1 + a_T S_z\sin\mathcal{C}\omex t\right)\,,
    \label{eq:Ez}
\end{equation}
and, similarly, $T_{t,\downarrow}\approx T(1 - a_T S_z\sin\mathcal{C}\omex t)$, with $a_T\approx \chi\Delta\Omega$.
Thus, we find that the final transverse temperature should correlate with the measured $S_z$ for $0<t<\pi/(\mathcal{C}\omex)$.
Eq.~\ref{eq:Ez} also predicts that the fluctuation $\Delta T_{t,\uparrow(\downarrow)}$ should have a time evolution similar to that of the amplification factor (Eq.~\ref{eq:alpha_model}), given the quantum fluctuations $\Delta S_z = \sqrt{N}/2$ of the initial state. Specifically, ignoring other fluctuations in temperature,
\begin{equation}
    \Delta T_{t,\uparrow}\approx \frac{\alpha_T \sqrt{N}}{2}\sin\mathcal{C}\omex t \propto \alpha(t)-1\, .
\end{equation}
This correlation is clearly demonstrated in Fig.~\ref{fig:temp_corr}(b).

\subsection{\label{app:nsim} Numerical simulation}
To better understand the amplification effect including lateral collisions and residual dephasing from the magnetic trap, we perform numerical simulations of the spin dynamics using a semiclassical kinetic equation for the spin vector $\bf{s}$ in the space of motional energies $\bf{E}=\lbrace E_x,E_y,E_z\rbrace$ \cite{Piechon2009,Deutsch2010}:
\begin{align}
&\partial_t\bf{s}(\bf{E},t) + \gamma_c[\bf{s}(\bf{E},t)- \bar{\bf{s}}] \nonumber \\
&= \left[ \delta\omega_a(\bm{r}(t),t)\bf{e}_z + \omex\int_0^\infty \dd\bf{E}'\beta^3e^{-\beta \bf{E}'}K(\bf{E},\bf{E}')\bf{s}(\bf{E}',t) \right] \nonumber\\
& \quad\times\bf{s}(\bf{E},t)
\label{eq:ISRE_kin}
\end{align}
where $\bar{\bf{s}}\equiv \int_0^\infty d\bf{E}\beta^3e^{-\beta \bf{E}}\bf{s}(\bf{E})$
describes the average spin. Integration is done on all three energies. $\bf{e}_z$ is the unit vector $\hat{z}$ in the Bloch sphere, generating spin precession at rate $\delta\omega_a(\bm{r},t)$ which includes three dephasing sources: ac Stark shift induced by the cavity probe (see Eq.~\ref{eq:cav_coupling}), shifts due to the magnetic trap and mean-field collisions \cite{Szmuk2015}. We include the spatial dependence of $\delta\omega_a$ to account for imperfections in the trap oscillation averaging (cf.~Eq.~\ref{eq:inhomo}).
The spin interaction depends on $\omega_\rm{ex}$ as well as the spin ``mean field'', and is long ranged in energy space (the Knudsen regime), described by the kernel $K(\bf{E},\bf{E'})$ which we approximate by $K(\bf{E},\bf{E}')\approx 1$ \cite{Deutsch2010,Pegahan2018} (this approximation slightly augments the exchange rate).
The lateral collision rate $\gamma_c$ is incorporated as a relaxation toward the mean spin.

To perform numerical simulations, we randomly sample the position and momentum of approximately $10^4$ atoms in a thermal distribution. The coordinates $\bm{r}(t)$ evolve as in pure harmonic oscillation. The atoms then have well-defined energies along each axis.
The cavity shift at each time step is calculated as $\sum_i\Omega(\bm{r}_i(t)) s_{z,i}(t)$ according to Eq.~\ref{eq:cav_coupling} with the $s_z$ component of each atom.
In order to simulate the amplification effect which amplifies quantum fluctuations in $S_z$, we start with all atoms having a common $s_z$ component that deviates from 0 (a classical approximation to the result of a QND cavity measurement). From the subsequently calculated cavity shift over time, we can obtain $M_1$ and $M_2$ hence their ratio $\alpha$.

In the simulation, we include the rea-time sequence of $M_1$ (two probes and composite $\pi$ pulse). However, we need to introduce decoherence (at single-spin level) to match the experimentally measured contrast (Fig.~\ref{fig:sqz_t}, inset); otherwise the spin-rotation rate would be overestimated. The theoretical lateral collision rate $\gamma_c$ alone is not sufficient to reproduce the strong damping in the measured $\alpha$. We introduce another phenomenological damping rate from the observed damping of center-of-mass oscillations ($\gamma_\rm{com}=0.45\,\rm{s}^{-1}$), which is mostly caused by the cavity-locking light (lateral collision would not damp the center-of-mass motion). $\gamma_c$ in Eq.~\ref{eq:ISRE_kin} is replaced by $\gamma_c+\gamma_\rm{com}$ in the simulation.

\bibliography{paper}

\providecommand{\noopsort}[1]{}\providecommand{\singleletter}[1]{#1}%
\begin{thebibliography}{50}%
\makeatletter
\providecommand \@ifxundefined [1]{%
 \@ifx{#1\undefined}
}%
\providecommand \@ifnum [1]{%
 \ifnum #1\expandafter \@firstoftwo
 \else \expandafter \@secondoftwo
 \fi
}%
\providecommand \@ifx [1]{%
 \ifx #1\expandafter \@firstoftwo
 \else \expandafter \@secondoftwo
 \fi
}%
\providecommand \natexlab [1]{#1}%
\providecommand \enquote  [1]{``#1''}%
\providecommand \bibnamefont  [1]{#1}%
\providecommand \bibfnamefont [1]{#1}%
\providecommand \citenamefont [1]{#1}%
\providecommand \href@noop [0]{\@secondoftwo}%
\providecommand \href [0]{\begingroup \@sanitize@url \@href}%
\providecommand \@href[1]{\@@startlink{#1}\@@href}%
\providecommand \@@href[1]{\endgroup#1\@@endlink}%
\providecommand \@sanitize@url [0]{\catcode `\\12\catcode `\$12\catcode
  `\&12\catcode `\#12\catcode `\^12\catcode `\_12\catcode `\%12\relax}%
\providecommand \@@startlink[1]{}%
\providecommand \@@endlink[0]{}%
\providecommand \url  [0]{\begingroup\@sanitize@url \@url }%
\providecommand \@url [1]{\endgroup\@href {#1}{\urlprefix }}%
\providecommand \urlprefix  [0]{URL }%
\providecommand \Eprint [0]{\href }%
\providecommand \doibase [0]{https://doi.org/}%
\providecommand \selectlanguage [0]{\@gobble}%
\providecommand \bibinfo  [0]{\@secondoftwo}%
\providecommand \bibfield  [0]{\@secondoftwo}%
\providecommand \translation [1]{[#1]}%
\providecommand \BibitemOpen [0]{}%
\providecommand \bibitemStop [0]{}%
\providecommand \bibitemNoStop [0]{.\EOS\space}%
\providecommand \EOS [0]{\spacefactor3000\relax}%
\providecommand \BibitemShut  [1]{\csname bibitem#1\endcsname}%
\let\auto@bib@innerbib\@empty
\bibitem [{\citenamefont {Kitagawa}\ and\ \citenamefont
  {Ueda}(1993)}]{Kitagawa1993}%
  \BibitemOpen
  \bibfield  {author} {\bibinfo {author} {\bibfnamefont {M.}~\bibnamefont
  {Kitagawa}}\ and\ \bibinfo {author} {\bibfnamefont {M.}~\bibnamefont
  {Ueda}},\ }\bibfield  {title} {\bibinfo {title} {Squeezed spin states},\
  }\href {https://doi.org/10.1103/physreva.47.5138} {\bibfield  {journal}
  {\bibinfo  {journal} {Physical Review A}\ }\textbf {\bibinfo {volume} {47}},\
  \bibinfo {pages} {5138} (\bibinfo {year} {1993})}\BibitemShut {NoStop}%
\bibitem [{\citenamefont {Wineland}\ \emph {et~al.}(1994)\citenamefont
  {Wineland}, \citenamefont {Bollinger}, \citenamefont {Itano},\ and\
  \citenamefont {Heinzen}}]{Wineland1994}%
  \BibitemOpen
  \bibfield  {author} {\bibinfo {author} {\bibfnamefont {D.~J.}\ \bibnamefont
  {Wineland}}, \bibinfo {author} {\bibfnamefont {J.~J.}\ \bibnamefont
  {Bollinger}}, \bibinfo {author} {\bibfnamefont {W.~M.}\ \bibnamefont
  {Itano}},\ and\ \bibinfo {author} {\bibfnamefont {D.~J.}\ \bibnamefont
  {Heinzen}},\ }\bibfield  {title} {\bibinfo {title} {Squeezed atomic states
  and projection noise in spectroscopy},\ }\href
  {https://doi.org/10.1103/physreva.50.67} {\bibfield  {journal} {\bibinfo
  {journal} {Physical Review A}\ }\textbf {\bibinfo {volume} {50}},\ \bibinfo
  {pages} {67} (\bibinfo {year} {1994})}\BibitemShut {NoStop}%
\bibitem [{\citenamefont {Ma}\ \emph {et~al.}(2011)\citenamefont {Ma},
  \citenamefont {Wang}, \citenamefont {Sun},\ and\ \citenamefont
  {Nori}}]{Ma2011}%
  \BibitemOpen
  \bibfield  {author} {\bibinfo {author} {\bibfnamefont {J.}~\bibnamefont
  {Ma}}, \bibinfo {author} {\bibfnamefont {X.}~\bibnamefont {Wang}}, \bibinfo
  {author} {\bibfnamefont {C.~P.}\ \bibnamefont {Sun}},\ and\ \bibinfo {author}
  {\bibfnamefont {F.}~\bibnamefont {Nori}},\ }\bibfield  {title} {\bibinfo
  {title} {Quantum spin squeezing},\ }\href
  {https://doi.org/10.1016/j.physrep.2011.08.003} {\bibfield  {journal}
  {\bibinfo  {journal} {Physics Reports}\ }\textbf {\bibinfo {volume} {509}},\
  \bibinfo {pages} {89} (\bibinfo {year} {2011})}\BibitemShut {NoStop}%
\bibitem [{\citenamefont {Pezz{\`{e}}}\ \emph {et~al.}(2018)\citenamefont
  {Pezz{\`{e}}}, \citenamefont {Smerzi}, \citenamefont {Oberthaler},
  \citenamefont {Schmied},\ and\ \citenamefont {Treutlein}}]{Pezze2018}%
  \BibitemOpen
  \bibfield  {author} {\bibinfo {author} {\bibfnamefont {L.}~\bibnamefont
  {Pezz{\`{e}}}}, \bibinfo {author} {\bibfnamefont {A.}~\bibnamefont {Smerzi}},
  \bibinfo {author} {\bibfnamefont {M.~K.}\ \bibnamefont {Oberthaler}},
  \bibinfo {author} {\bibfnamefont {R.}~\bibnamefont {Schmied}},\ and\ \bibinfo
  {author} {\bibfnamefont {P.}~\bibnamefont {Treutlein}},\ }\bibfield  {title}
  {\bibinfo {title} {Quantum metrology with nonclassical states of atomic
  ensembles},\ }\href {https://doi.org/10.1103/revmodphys.90.035005} {\bibfield
   {journal} {\bibinfo  {journal} {Reviews of Modern Physics}\ }\textbf
  {\bibinfo {volume} {90}},\ \bibinfo {pages} {035005} (\bibinfo {year}
  {2018})}\BibitemShut {NoStop}%
\bibitem [{\citenamefont {Santarelli}\ \emph {et~al.}(1999)\citenamefont
  {Santarelli}, \citenamefont {Laurent}, \citenamefont {Lemonde}, \citenamefont
  {Clairon}, \citenamefont {Mann}, \citenamefont {Chang}, \citenamefont
  {Luiten},\ and\ \citenamefont {Salomon}}]{Santarelli1999}%
  \BibitemOpen
  \bibfield  {author} {\bibinfo {author} {\bibfnamefont {G.}~\bibnamefont
  {Santarelli}}, \bibinfo {author} {\bibfnamefont {P.}~\bibnamefont {Laurent}},
  \bibinfo {author} {\bibfnamefont {P.}~\bibnamefont {Lemonde}}, \bibinfo
  {author} {\bibfnamefont {A.}~\bibnamefont {Clairon}}, \bibinfo {author}
  {\bibfnamefont {A.~G.}\ \bibnamefont {Mann}}, \bibinfo {author}
  {\bibfnamefont {S.}~\bibnamefont {Chang}}, \bibinfo {author} {\bibfnamefont
  {A.~N.}\ \bibnamefont {Luiten}},\ and\ \bibinfo {author} {\bibfnamefont
  {C.}~\bibnamefont {Salomon}},\ }\bibfield  {title} {\bibinfo {title} {Quantum
  projection noise in an atomic fountain: A high stability cesium frequency
  standard},\ }\href {https://doi.org/10.1103/PhysRevLett.82.4619} {\bibfield
  {journal} {\bibinfo  {journal} {Physical Review Letters}\ }\textbf {\bibinfo
  {volume} {82}},\ \bibinfo {pages} {4619} (\bibinfo {year}
  {1999})}\BibitemShut {NoStop}%
\bibitem [{\citenamefont {Gauguet}\ \emph {et~al.}(2009)\citenamefont
  {Gauguet}, \citenamefont {Canuel}, \citenamefont {L\'{e}v\`{e}que},
  \citenamefont {Chaibi},\ and\ \citenamefont {Landragin}}]{Gauguet2009}%
  \BibitemOpen
  \bibfield  {author} {\bibinfo {author} {\bibfnamefont {A.}~\bibnamefont
  {Gauguet}}, \bibinfo {author} {\bibfnamefont {B.}~\bibnamefont {Canuel}},
  \bibinfo {author} {\bibfnamefont {T.}~\bibnamefont {L\'{e}v\`{e}que}},
  \bibinfo {author} {\bibfnamefont {W.}~\bibnamefont {Chaibi}},\ and\ \bibinfo
  {author} {\bibfnamefont {A.}~\bibnamefont {Landragin}},\ }\bibfield  {title}
  {\bibinfo {title} {Characterization and limits of a cold-atom {Sagnac}
  interferometer},\ }\href {https://doi.org/10.1103/PhysRevA.80.063604}
  {\bibfield  {journal} {\bibinfo  {journal} {Physical Review A}\ }\textbf
  {\bibinfo {volume} {80}},\ \bibinfo {pages} {063604} (\bibinfo {year}
  {2009})}\BibitemShut {NoStop}%
\bibitem [{\citenamefont {Rosi}\ \emph {et~al.}(2014)\citenamefont {Rosi},
  \citenamefont {Sorrentino}, \citenamefont {Cacciapuoti}, \citenamefont
  {Prevedelli},\ and\ \citenamefont {Tino}}]{Rosi2014}%
  \BibitemOpen
  \bibfield  {author} {\bibinfo {author} {\bibfnamefont {G.}~\bibnamefont
  {Rosi}}, \bibinfo {author} {\bibfnamefont {F.}~\bibnamefont {Sorrentino}},
  \bibinfo {author} {\bibfnamefont {L.}~\bibnamefont {Cacciapuoti}}, \bibinfo
  {author} {\bibfnamefont {M.}~\bibnamefont {Prevedelli}},\ and\ \bibinfo
  {author} {\bibfnamefont {G.~M.}\ \bibnamefont {Tino}},\ }\bibfield  {title}
  {\bibinfo {title} {Precision measurement of the newtonian gravitational
  constant using cold atoms},\ }\href {https://doi.org/10.1038/nature13433}
  {\bibfield  {journal} {\bibinfo  {journal} {Nature}\ }\textbf {\bibinfo
  {volume} {510}},\ \bibinfo {pages} {518} (\bibinfo {year}
  {2014})}\BibitemShut {NoStop}%
\bibitem [{\citenamefont {Oelker}\ \emph {et~al.}(2019)\citenamefont {Oelker},
  \citenamefont {Hutson}, \citenamefont {Kennedy}, \citenamefont {Sonderhouse},
  \citenamefont {Bothwell}, \citenamefont {Goban}, \citenamefont {Kedar},
  \citenamefont {Sanner}, \citenamefont {Robinson}, \citenamefont {Marti},
  \citenamefont {Matei}, \citenamefont {Legero}, \citenamefont {Giunta},
  \citenamefont {Holzwarth}, \citenamefont {Riehle}, \citenamefont {Sterr},\
  and\ \citenamefont {Ye}}]{Oelker2019}%
  \BibitemOpen
  \bibfield  {author} {\bibinfo {author} {\bibfnamefont {E.}~\bibnamefont
  {Oelker}}, \bibinfo {author} {\bibfnamefont {R.~B.}\ \bibnamefont {Hutson}},
  \bibinfo {author} {\bibfnamefont {C.~J.}\ \bibnamefont {Kennedy}}, \bibinfo
  {author} {\bibfnamefont {L.}~\bibnamefont {Sonderhouse}}, \bibinfo {author}
  {\bibfnamefont {T.}~\bibnamefont {Bothwell}}, \bibinfo {author}
  {\bibfnamefont {A.}~\bibnamefont {Goban}}, \bibinfo {author} {\bibfnamefont
  {D.}~\bibnamefont {Kedar}}, \bibinfo {author} {\bibfnamefont
  {C.}~\bibnamefont {Sanner}}, \bibinfo {author} {\bibfnamefont {J.~M.}\
  \bibnamefont {Robinson}}, \bibinfo {author} {\bibfnamefont {G.~E.}\
  \bibnamefont {Marti}}, \bibinfo {author} {\bibfnamefont {D.~G.}\ \bibnamefont
  {Matei}}, \bibinfo {author} {\bibfnamefont {T.}~\bibnamefont {Legero}},
  \bibinfo {author} {\bibfnamefont {M.}~\bibnamefont {Giunta}}, \bibinfo
  {author} {\bibfnamefont {R.}~\bibnamefont {Holzwarth}}, \bibinfo {author}
  {\bibfnamefont {F.}~\bibnamefont {Riehle}}, \bibinfo {author} {\bibfnamefont
  {U.}~\bibnamefont {Sterr}},\ and\ \bibinfo {author} {\bibfnamefont
  {J.}~\bibnamefont {Ye}},\ }\bibfield  {title} {\bibinfo {title}
  {Demonstration of $4.8\times 10^{-17}$ stability at 1\,s for two independent
  optical clocks},\ }\href {https://doi.org/10.1038/s41566-019-0493-4}
  {\bibfield  {journal} {\bibinfo  {journal} {Nature Photonics}\ }\textbf
  {\bibinfo {volume} {13}},\ \bibinfo {pages} {714} (\bibinfo {year}
  {2019})}\BibitemShut {NoStop}%
\bibitem [{\citenamefont {Wasilewski}\ \emph {et~al.}(2010)\citenamefont
  {Wasilewski}, \citenamefont {Jensen}, \citenamefont {Krauter}, \citenamefont
  {Renema}, \citenamefont {Balabas},\ and\ \citenamefont
  {Polzik}}]{Wasilewski2010}%
  \BibitemOpen
  \bibfield  {author} {\bibinfo {author} {\bibfnamefont {W.}~\bibnamefont
  {Wasilewski}}, \bibinfo {author} {\bibfnamefont {K.}~\bibnamefont {Jensen}},
  \bibinfo {author} {\bibfnamefont {H.}~\bibnamefont {Krauter}}, \bibinfo
  {author} {\bibfnamefont {J.~J.}\ \bibnamefont {Renema}}, \bibinfo {author}
  {\bibfnamefont {M.~V.}\ \bibnamefont {Balabas}},\ and\ \bibinfo {author}
  {\bibfnamefont {E.~S.}\ \bibnamefont {Polzik}},\ }\bibfield  {title}
  {\bibinfo {title} {Quantum noise limited and entanglement-assisted
  magnetometry},\ }\href {https://doi.org/10.1103/physrevlett.104.133601}
  {\bibfield  {journal} {\bibinfo  {journal} {Physical Review Letters}\
  }\textbf {\bibinfo {volume} {104}},\ \bibinfo {pages} {133601} (\bibinfo
  {year} {2010})}\BibitemShut {NoStop}%
\bibitem [{\citenamefont {Appel}\ \emph {et~al.}(2009)\citenamefont {Appel},
  \citenamefont {Windpassinger}, \citenamefont {Oblak}, \citenamefont {Hoff},
  \citenamefont {Kjaergaard},\ and\ \citenamefont {Polzik}}]{Appel2009}%
  \BibitemOpen
  \bibfield  {author} {\bibinfo {author} {\bibfnamefont {J.}~\bibnamefont
  {Appel}}, \bibinfo {author} {\bibfnamefont {P.~J.}\ \bibnamefont
  {Windpassinger}}, \bibinfo {author} {\bibfnamefont {D.}~\bibnamefont
  {Oblak}}, \bibinfo {author} {\bibfnamefont {U.~B.}\ \bibnamefont {Hoff}},
  \bibinfo {author} {\bibfnamefont {N.}~\bibnamefont {Kjaergaard}},\ and\
  \bibinfo {author} {\bibfnamefont {E.~S.}\ \bibnamefont {Polzik}},\ }\bibfield
   {title} {\bibinfo {title} {Mesoscopic atomic entanglement for precision
  measurements beyond the standard quantum limit},\ }\href
  {https://doi.org/10.1073/pnas.0901550106} {\bibfield  {journal} {\bibinfo
  {journal} {Proceedings of the National Academy of Sciences}\ }\textbf
  {\bibinfo {volume} {106}},\ \bibinfo {pages} {10960} (\bibinfo {year}
  {2009})}\BibitemShut {NoStop}%
\bibitem [{\citenamefont {Leroux}\ \emph
  {et~al.}(2010{\natexlab{a}})\citenamefont {Leroux}, \citenamefont
  {Schleier-Smith},\ and\ \citenamefont {Vuleti{\'{c}}}}]{Leroux2010}%
  \BibitemOpen
  \bibfield  {author} {\bibinfo {author} {\bibfnamefont {I.~D.}\ \bibnamefont
  {Leroux}}, \bibinfo {author} {\bibfnamefont {M.~H.}\ \bibnamefont
  {Schleier-Smith}},\ and\ \bibinfo {author} {\bibfnamefont {V.}~\bibnamefont
  {Vuleti{\'{c}}}},\ }\bibfield  {title} {\bibinfo {title} {Implementation of
  cavity squeezing of a collective atomic spin},\ }\href
  {https://doi.org/10.1103/physrevlett.104.073602} {\bibfield  {journal}
  {\bibinfo  {journal} {Physical Review Letters}\ }\textbf {\bibinfo {volume}
  {104}},\ \bibinfo {pages} {073602} (\bibinfo {year}
  {2010}{\natexlab{a}})}\BibitemShut {NoStop}%
\bibitem [{\citenamefont {Schleier-Smith}\ \emph {et~al.}(2010)\citenamefont
  {Schleier-Smith}, \citenamefont {Leroux},\ and\ \citenamefont
  {Vuleti{\'{c}}}}]{Schleier-Smith2010a}%
  \BibitemOpen
  \bibfield  {author} {\bibinfo {author} {\bibfnamefont {M.~H.}\ \bibnamefont
  {Schleier-Smith}}, \bibinfo {author} {\bibfnamefont {I.~D.}\ \bibnamefont
  {Leroux}},\ and\ \bibinfo {author} {\bibfnamefont {V.}~\bibnamefont
  {Vuleti{\'{c}}}},\ }\bibfield  {title} {\bibinfo {title} {States of an
  ensemble of two-level atoms with reduced quantum uncertainty},\ }\href
  {https://doi.org/10.1103/physrevlett.104.073604} {\bibfield  {journal}
  {\bibinfo  {journal} {Physical Review Letters}\ }\textbf {\bibinfo {volume}
  {104}},\ \bibinfo {pages} {073604} (\bibinfo {year} {2010})}\BibitemShut
  {NoStop}%
\bibitem [{\citenamefont {Riedel}\ \emph {et~al.}(2010)\citenamefont {Riedel},
  \citenamefont {B\"ohi}, \citenamefont {Li}, \citenamefont {H\"ansch},
  \citenamefont {Sinatra},\ and\ \citenamefont {Treutlein}}]{Riedel2010}%
  \BibitemOpen
  \bibfield  {author} {\bibinfo {author} {\bibfnamefont {M.~F.}\ \bibnamefont
  {Riedel}}, \bibinfo {author} {\bibfnamefont {P.}~\bibnamefont {B\"ohi}},
  \bibinfo {author} {\bibfnamefont {Y.}~\bibnamefont {Li}}, \bibinfo {author}
  {\bibfnamefont {T.~W.}\ \bibnamefont {H\"ansch}}, \bibinfo {author}
  {\bibfnamefont {A.}~\bibnamefont {Sinatra}},\ and\ \bibinfo {author}
  {\bibfnamefont {P.}~\bibnamefont {Treutlein}},\ }\bibfield  {title} {\bibinfo
  {title} {Atom-chip-based generation of entanglement for quantum metrology},\
  }\href {https://doi.org/10.1038/nature08988} {\bibfield  {journal} {\bibinfo
  {journal} {Nature}\ }\textbf {\bibinfo {volume} {464}},\ \bibinfo {pages}
  {1170} (\bibinfo {year} {2010})}\BibitemShut {NoStop}%
\bibitem [{\citenamefont {Gross}\ \emph {et~al.}(2010)\citenamefont {Gross},
  \citenamefont {Zibold}, \citenamefont {Nicklas}, \citenamefont {Est{\`e}ve},\
  and\ \citenamefont {Oberthaler}}]{Gross2010}%
  \BibitemOpen
  \bibfield  {author} {\bibinfo {author} {\bibfnamefont {C.}~\bibnamefont
  {Gross}}, \bibinfo {author} {\bibfnamefont {T.}~\bibnamefont {Zibold}},
  \bibinfo {author} {\bibfnamefont {E.}~\bibnamefont {Nicklas}}, \bibinfo
  {author} {\bibfnamefont {J.}~\bibnamefont {Est{\`e}ve}},\ and\ \bibinfo
  {author} {\bibfnamefont {M.~K.}\ \bibnamefont {Oberthaler}},\ }\bibfield
  {title} {\bibinfo {title} {Nonlinear atom interferometer surpasses classical
  precision limit},\ }\href {https://doi.org/10.1038/nature08919} {\bibfield
  {journal} {\bibinfo  {journal} {Nature}\ }\textbf {\bibinfo {volume} {464}},\
  \bibinfo {pages} {1165} (\bibinfo {year} {2010})}\BibitemShut {NoStop}%
\bibitem [{\citenamefont {Bohnet}\ \emph {et~al.}(2014)\citenamefont {Bohnet},
  \citenamefont {Cox}, \citenamefont {Norcia}, \citenamefont {Weiner},
  \citenamefont {Chen},\ and\ \citenamefont {Thompson}}]{Bohnet2014}%
  \BibitemOpen
  \bibfield  {author} {\bibinfo {author} {\bibfnamefont {J.~G.}\ \bibnamefont
  {Bohnet}}, \bibinfo {author} {\bibfnamefont {K.~C.}\ \bibnamefont {Cox}},
  \bibinfo {author} {\bibfnamefont {M.~A.}\ \bibnamefont {Norcia}}, \bibinfo
  {author} {\bibfnamefont {J.~M.}\ \bibnamefont {Weiner}}, \bibinfo {author}
  {\bibfnamefont {Z.}~\bibnamefont {Chen}},\ and\ \bibinfo {author}
  {\bibfnamefont {J.~K.}\ \bibnamefont {Thompson}},\ }\bibfield  {title}
  {\bibinfo {title} {Reduced spin measurement back-action for a phase
  sensitivity ten times beyond the standard quantum limit},\ }\href
  {https://doi.org/10.1038/nphoton.2014.151} {\bibfield  {journal} {\bibinfo
  {journal} {Nature Photonics}\ }\textbf {\bibinfo {volume} {8}},\ \bibinfo
  {pages} {731} (\bibinfo {year} {2014})}\BibitemShut {NoStop}%
\bibitem [{\citenamefont {Cox}\ \emph {et~al.}(2016)\citenamefont {Cox},
  \citenamefont {Greve}, \citenamefont {Weiner},\ and\ \citenamefont
  {Thompson}}]{Cox2016}%
  \BibitemOpen
  \bibfield  {author} {\bibinfo {author} {\bibfnamefont {K.~C.}\ \bibnamefont
  {Cox}}, \bibinfo {author} {\bibfnamefont {G.~P.}\ \bibnamefont {Greve}},
  \bibinfo {author} {\bibfnamefont {J.~M.}\ \bibnamefont {Weiner}},\ and\
  \bibinfo {author} {\bibfnamefont {J.~K.}\ \bibnamefont {Thompson}},\
  }\bibfield  {title} {\bibinfo {title} {Deterministic squeezed states with
  collective measurements and feedback},\ }\href
  {https://doi.org/10.1103/physrevlett.116.093602} {\bibfield  {journal}
  {\bibinfo  {journal} {Physical Review Letters}\ }\textbf {\bibinfo {volume}
  {116}},\ \bibinfo {pages} {093602} (\bibinfo {year} {2016})}\BibitemShut
  {NoStop}%
\bibitem [{\citenamefont {Bohnet}\ \emph {et~al.}(2016)\citenamefont {Bohnet},
  \citenamefont {Sawyer}, \citenamefont {Britton}, \citenamefont {Wall},
  \citenamefont {Rey}, \citenamefont {Foss-Feig},\ and\ \citenamefont
  {Bollinger}}]{Bohnet2016}%
  \BibitemOpen
  \bibfield  {author} {\bibinfo {author} {\bibfnamefont {J.~G.}\ \bibnamefont
  {Bohnet}}, \bibinfo {author} {\bibfnamefont {B.~C.}\ \bibnamefont {Sawyer}},
  \bibinfo {author} {\bibfnamefont {J.~W.}\ \bibnamefont {Britton}}, \bibinfo
  {author} {\bibfnamefont {M.~L.}\ \bibnamefont {Wall}}, \bibinfo {author}
  {\bibfnamefont {A.~M.}\ \bibnamefont {Rey}}, \bibinfo {author} {\bibfnamefont
  {M.}~\bibnamefont {Foss-Feig}},\ and\ \bibinfo {author} {\bibfnamefont
  {J.~J.}\ \bibnamefont {Bollinger}},\ }\bibfield  {title} {\bibinfo {title}
  {Quantum spin dynamics and entanglement generation with hundreds of trapped
  ions},\ }\href {https://doi.org/10.1126/science.aad9958} {\bibfield
  {journal} {\bibinfo  {journal} {Science}\ }\textbf {\bibinfo {volume}
  {352}},\ \bibinfo {pages} {1297} (\bibinfo {year} {2016})}\BibitemShut
  {NoStop}%
\bibitem [{\citenamefont {Hosten}\ \emph
  {et~al.}(2016{\natexlab{a}})\citenamefont {Hosten}, \citenamefont {Engelsen},
  \citenamefont {Krishnakumar},\ and\ \citenamefont {Kasevich}}]{Hosten2016}%
  \BibitemOpen
  \bibfield  {author} {\bibinfo {author} {\bibfnamefont {O.}~\bibnamefont
  {Hosten}}, \bibinfo {author} {\bibfnamefont {N.~J.}\ \bibnamefont
  {Engelsen}}, \bibinfo {author} {\bibfnamefont {R.}~\bibnamefont
  {Krishnakumar}},\ and\ \bibinfo {author} {\bibfnamefont {M.~A.}\ \bibnamefont
  {Kasevich}},\ }\bibfield  {title} {\bibinfo {title} {Measurement noise 100
  times lower than the quantum-projection limit using entangled atoms},\ }\href
  {https://doi.org/10.1038/nature16176} {\bibfield  {journal} {\bibinfo
  {journal} {Nature}\ }\textbf {\bibinfo {volume} {529}},\ \bibinfo {pages}
  {505} (\bibinfo {year} {2016}{\natexlab{a}})}\BibitemShut {NoStop}%
\bibitem [{\citenamefont {Leroux}\ \emph
  {et~al.}(2010{\natexlab{b}})\citenamefont {Leroux}, \citenamefont
  {Schleier-Smith},\ and\ \citenamefont {Vuleti{\'{c}}}}]{Leroux2010a}%
  \BibitemOpen
  \bibfield  {author} {\bibinfo {author} {\bibfnamefont {I.~D.}\ \bibnamefont
  {Leroux}}, \bibinfo {author} {\bibfnamefont {M.~H.}\ \bibnamefont
  {Schleier-Smith}},\ and\ \bibinfo {author} {\bibfnamefont {V.}~\bibnamefont
  {Vuleti{\'{c}}}},\ }\bibfield  {title} {\bibinfo {title}
  {Orientation-dependent entanglement lifetime in a squeezed atomic clock},\
  }\href {https://doi.org/10.1103/physrevlett.104.250801} {\bibfield  {journal}
  {\bibinfo  {journal} {Physical Review Letters}\ }\textbf {\bibinfo {volume}
  {104}},\ \bibinfo {pages} {250801} (\bibinfo {year}
  {2010}{\natexlab{b}})}\BibitemShut {NoStop}%
\bibitem [{\citenamefont {Sewell}\ \emph {et~al.}(2012)\citenamefont {Sewell},
  \citenamefont {Koschorreck}, \citenamefont {Napolitano}, \citenamefont
  {Dubost}, \citenamefont {Behbood},\ and\ \citenamefont
  {Mitchell}}]{Sewell2012}%
  \BibitemOpen
  \bibfield  {author} {\bibinfo {author} {\bibfnamefont {R.~J.}\ \bibnamefont
  {Sewell}}, \bibinfo {author} {\bibfnamefont {M.}~\bibnamefont {Koschorreck}},
  \bibinfo {author} {\bibfnamefont {M.}~\bibnamefont {Napolitano}}, \bibinfo
  {author} {\bibfnamefont {B.}~\bibnamefont {Dubost}}, \bibinfo {author}
  {\bibfnamefont {N.}~\bibnamefont {Behbood}},\ and\ \bibinfo {author}
  {\bibfnamefont {M.~W.}\ \bibnamefont {Mitchell}},\ }\bibfield  {title}
  {\bibinfo {title} {Magnetic sensitivity beyond the projection noise limit by
  spin squeezing},\ }\href {https://doi.org/10.1103/physrevlett.109.253605}
  {\bibfield  {journal} {\bibinfo  {journal} {Physical Review Letters}\
  }\textbf {\bibinfo {volume} {109}},\ \bibinfo {pages} {253605} (\bibinfo
  {year} {2012})}\BibitemShut {NoStop}%
\bibitem [{\citenamefont {Bao}\ \emph {et~al.}(2020)\citenamefont {Bao},
  \citenamefont {Duan}, \citenamefont {Jin}, \citenamefont {Lu}, \citenamefont
  {Li}, \citenamefont {Qu}, \citenamefont {Wang}, \citenamefont {Novikova},
  \citenamefont {Mikhailov}, \citenamefont {Zhao}, \citenamefont {M{\o}lmer},
  \citenamefont {Shen},\ and\ \citenamefont {Xiao}}]{Bao2020}%
  \BibitemOpen
  \bibfield  {author} {\bibinfo {author} {\bibfnamefont {H.}~\bibnamefont
  {Bao}}, \bibinfo {author} {\bibfnamefont {J.}~\bibnamefont {Duan}}, \bibinfo
  {author} {\bibfnamefont {S.}~\bibnamefont {Jin}}, \bibinfo {author}
  {\bibfnamefont {X.}~\bibnamefont {Lu}}, \bibinfo {author} {\bibfnamefont
  {P.}~\bibnamefont {Li}}, \bibinfo {author} {\bibfnamefont {W.}~\bibnamefont
  {Qu}}, \bibinfo {author} {\bibfnamefont {M.}~\bibnamefont {Wang}}, \bibinfo
  {author} {\bibfnamefont {I.}~\bibnamefont {Novikova}}, \bibinfo {author}
  {\bibfnamefont {E.~E.}\ \bibnamefont {Mikhailov}}, \bibinfo {author}
  {\bibfnamefont {K.-F.}\ \bibnamefont {Zhao}}, \bibinfo {author}
  {\bibfnamefont {K.}~\bibnamefont {M{\o}lmer}}, \bibinfo {author}
  {\bibfnamefont {H.}~\bibnamefont {Shen}},\ and\ \bibinfo {author}
  {\bibfnamefont {Y.}~\bibnamefont {Xiao}},\ }\bibfield  {title} {\bibinfo
  {title} {Spin squeezing of 1011 atoms by prediction and retrodiction
  measurements},\ }\href {https://doi.org/10.1038/s41586-020-2243-7} {\bibfield
   {journal} {\bibinfo  {journal} {Nature}\ }\textbf {\bibinfo {volume}
  {581}},\ \bibinfo {pages} {159} (\bibinfo {year} {2020})}\BibitemShut
  {NoStop}%
\bibitem [{\citenamefont {Pedrozo-Pe{\~n}afiel}\ \emph
  {et~al.}(2020)\citenamefont {Pedrozo-Pe{\~n}afiel}, \citenamefont {Colombo},
  \citenamefont {Shu}, \citenamefont {Adiyatullin}, \citenamefont {Li},
  \citenamefont {Mendez}, \citenamefont {Braverman}, \citenamefont {Kawasaki},
  \citenamefont {Akamatsu}, \citenamefont {Xiao},\ and\ \citenamefont
  {Vuleti\'c}}]{Pedrozo2020}%
  \BibitemOpen
  \bibfield  {author} {\bibinfo {author} {\bibfnamefont {E.}~\bibnamefont
  {Pedrozo-Pe{\~n}afiel}}, \bibinfo {author} {\bibfnamefont {S.}~\bibnamefont
  {Colombo}}, \bibinfo {author} {\bibfnamefont {C.}~\bibnamefont {Shu}},
  \bibinfo {author} {\bibfnamefont {A.~F.}\ \bibnamefont {Adiyatullin}},
  \bibinfo {author} {\bibfnamefont {Z.}~\bibnamefont {Li}}, \bibinfo {author}
  {\bibfnamefont {E.}~\bibnamefont {Mendez}}, \bibinfo {author} {\bibfnamefont
  {B.}~\bibnamefont {Braverman}}, \bibinfo {author} {\bibfnamefont
  {A.}~\bibnamefont {Kawasaki}}, \bibinfo {author} {\bibfnamefont
  {D.}~\bibnamefont {Akamatsu}}, \bibinfo {author} {\bibfnamefont
  {Y.}~\bibnamefont {Xiao}},\ and\ \bibinfo {author} {\bibfnamefont
  {V.}~\bibnamefont {Vuleti\'c}},\ }\bibfield  {title} {\bibinfo {title}
  {Entanglement on an optical atomic-clock transition},\ }\href
  {https://doi.org/10.1038/s41586-020-3006-1} {\bibfield  {journal} {\bibinfo
  {journal} {Nature}\ }\textbf {\bibinfo {volume} {588}},\ \bibinfo {pages}
  {414} (\bibinfo {year} {2020})}\BibitemShut {NoStop}%
\bibitem [{\citenamefont {Robinson}\ \emph {et~al.}(2022)\citenamefont
  {Robinson}, \citenamefont {Miklos}, \citenamefont {Tso}, \citenamefont
  {Kennedy}, \citenamefont {Bothwell}, \citenamefont {Kedar}, \citenamefont
  {Thompson},\ and\ \citenamefont {Ye}}]{robinson_direct_2022}%
  \BibitemOpen
  \bibfield  {author} {\bibinfo {author} {\bibfnamefont {J.~M.}\ \bibnamefont
  {Robinson}}, \bibinfo {author} {\bibfnamefont {M.}~\bibnamefont {Miklos}},
  \bibinfo {author} {\bibfnamefont {Y.~M.}\ \bibnamefont {Tso}}, \bibinfo
  {author} {\bibfnamefont {C.~J.}\ \bibnamefont {Kennedy}}, \bibinfo {author}
  {\bibfnamefont {T.}~\bibnamefont {Bothwell}}, \bibinfo {author}
  {\bibfnamefont {D.}~\bibnamefont {Kedar}}, \bibinfo {author} {\bibfnamefont
  {J.~K.}\ \bibnamefont {Thompson}},\ and\ \bibinfo {author} {\bibfnamefont
  {J.}~\bibnamefont {Ye}},\ }\href {https://doi.org/10.48550/arXiv.2211.08621}
  {\bibinfo {title} {Direct comparison of two spin squeezed optical clocks
  below the quantum projection noise limit}} (\bibinfo {year} {2022}),\
  \bibinfo {note} {arXiv:2211.08621 [physics, physics:quant-ph]}\BibitemShut
  {NoStop}%
\bibitem [{\citenamefont {Malia}\ \emph {et~al.}(2020)\citenamefont {Malia},
  \citenamefont {Mart{\'{\i}}nez-Rinc{\'{o}}n}, \citenamefont {Wu},
  \citenamefont {Hosten},\ and\ \citenamefont {Kasevich}}]{Malia2020}%
  \BibitemOpen
  \bibfield  {author} {\bibinfo {author} {\bibfnamefont {B.~K.}\ \bibnamefont
  {Malia}}, \bibinfo {author} {\bibfnamefont {J.}~\bibnamefont
  {Mart{\'{\i}}nez-Rinc{\'{o}}n}}, \bibinfo {author} {\bibfnamefont
  {Y.}~\bibnamefont {Wu}}, \bibinfo {author} {\bibfnamefont {O.}~\bibnamefont
  {Hosten}},\ and\ \bibinfo {author} {\bibfnamefont {M.~A.}\ \bibnamefont
  {Kasevich}},\ }\bibfield  {title} {\bibinfo {title} {Free space ramsey
  spectroscopy in rubidium with noise below the quantum projection limit},\
  }\href {https://doi.org/10.1103/physrevlett.125.043202} {\bibfield  {journal}
  {\bibinfo  {journal} {Physical Review Letters}\ }\textbf {\bibinfo {volume}
  {125}},\ \bibinfo {pages} {043202} (\bibinfo {year} {2020})}\BibitemShut
  {NoStop}%
\bibitem [{\citenamefont {Gu{\'e}na}\ \emph {et~al.}(2012)\citenamefont
  {Gu{\'e}na}, \citenamefont {Abgrall}, \citenamefont {Rovera}, \citenamefont
  {Laurent}, \citenamefont {Chupin}, \citenamefont {Lours}, \citenamefont
  {Santarelli}, \citenamefont {Rosenbusch}, \citenamefont {Tobar},
  \citenamefont {Li}, \citenamefont {Gibble}, \citenamefont {Clairon},\ and\
  \citenamefont {Bize}}]{Guena2012}%
  \BibitemOpen
  \bibfield  {author} {\bibinfo {author} {\bibfnamefont {J.}~\bibnamefont
  {Gu{\'e}na}}, \bibinfo {author} {\bibfnamefont {M.}~\bibnamefont {Abgrall}},
  \bibinfo {author} {\bibfnamefont {D.}~\bibnamefont {Rovera}}, \bibinfo
  {author} {\bibfnamefont {P.}~\bibnamefont {Laurent}}, \bibinfo {author}
  {\bibfnamefont {B.}~\bibnamefont {Chupin}}, \bibinfo {author} {\bibfnamefont
  {M.}~\bibnamefont {Lours}}, \bibinfo {author} {\bibfnamefont
  {G.}~\bibnamefont {Santarelli}}, \bibinfo {author} {\bibfnamefont
  {P.}~\bibnamefont {Rosenbusch}}, \bibinfo {author} {\bibfnamefont {M.~E.}\
  \bibnamefont {Tobar}}, \bibinfo {author} {\bibfnamefont {R.}~\bibnamefont
  {Li}}, \bibinfo {author} {\bibfnamefont {K.}~\bibnamefont {Gibble}}, \bibinfo
  {author} {\bibfnamefont {A.}~\bibnamefont {Clairon}},\ and\ \bibinfo {author}
  {\bibfnamefont {S.}~\bibnamefont {Bize}},\ }\bibfield  {title} {\bibinfo
  {title} {Progress in atomic fountains at {LNE}-{SYRTE}},\ }\href
  {https://doi.org/10.1109/TUFFC.2012.2208} {\bibfield  {journal} {\bibinfo
  {journal} {{IEEE} Transactions on Ultrasonics, Ferroelectrics and Frequency
  Control}\ }\textbf {\bibinfo {volume} {59}},\ \bibinfo {pages} {391}
  (\bibinfo {year} {2012})}\BibitemShut {NoStop}%
\bibitem [{\citenamefont {Ludlow}\ \emph {et~al.}(2015)\citenamefont {Ludlow},
  \citenamefont {Boyd}, \citenamefont {Ye}, \citenamefont {Peik},\ and\
  \citenamefont {Schmidt}}]{Ludlow2015}%
  \BibitemOpen
  \bibfield  {author} {\bibinfo {author} {\bibfnamefont {A.~D.}\ \bibnamefont
  {Ludlow}}, \bibinfo {author} {\bibfnamefont {M.~M.}\ \bibnamefont {Boyd}},
  \bibinfo {author} {\bibfnamefont {J.}~\bibnamefont {Ye}}, \bibinfo {author}
  {\bibfnamefont {E.}~\bibnamefont {Peik}},\ and\ \bibinfo {author}
  {\bibfnamefont {P.~O.}\ \bibnamefont {Schmidt}},\ }\bibfield  {title}
  {\bibinfo {title} {Optical atomic clocks},\ }\href
  {https://doi.org/10.1103/revmodphys.87.637} {\bibfield  {journal} {\bibinfo
  {journal} {Reviews of Modern Physics}\ }\textbf {\bibinfo {volume} {87}},\
  \bibinfo {pages} {637} (\bibinfo {year} {2015})}\BibitemShut {NoStop}%
\bibitem [{\citenamefont {Barrett}\ \emph {et~al.}(2016)\citenamefont
  {Barrett}, \citenamefont {Bertoldi},\ and\ \citenamefont
  {Bouyer}}]{Barrett2016}%
  \BibitemOpen
  \bibfield  {author} {\bibinfo {author} {\bibfnamefont {B.}~\bibnamefont
  {Barrett}}, \bibinfo {author} {\bibfnamefont {A.}~\bibnamefont {Bertoldi}},\
  and\ \bibinfo {author} {\bibfnamefont {P.}~\bibnamefont {Bouyer}},\
  }\bibfield  {title} {\bibinfo {title} {Inertial quantum sensors using light
  and matter},\ }\href {https://doi.org/10.1088/0031-8949/91/5/053006}
  {\bibfield  {journal} {\bibinfo  {journal} {Physica Scripta}\ }\textbf
  {\bibinfo {volume} {91}},\ \bibinfo {pages} {053006} (\bibinfo {year}
  {2016})}\BibitemShut {NoStop}%
\bibitem [{\citenamefont {Weyers}\ \emph {et~al.}(2018)\citenamefont {Weyers},
  \citenamefont {Gerginov}, \citenamefont {Kazda}, \citenamefont {Rahm},
  \citenamefont {Lipphardt}, \citenamefont {Dobrev},\ and\ \citenamefont
  {Gibble}}]{Weyers2018}%
  \BibitemOpen
  \bibfield  {author} {\bibinfo {author} {\bibfnamefont {S.}~\bibnamefont
  {Weyers}}, \bibinfo {author} {\bibfnamefont {V.}~\bibnamefont {Gerginov}},
  \bibinfo {author} {\bibfnamefont {M.}~\bibnamefont {Kazda}}, \bibinfo
  {author} {\bibfnamefont {J.}~\bibnamefont {Rahm}}, \bibinfo {author}
  {\bibfnamefont {B.}~\bibnamefont {Lipphardt}}, \bibinfo {author}
  {\bibfnamefont {G.}~\bibnamefont {Dobrev}},\ and\ \bibinfo {author}
  {\bibfnamefont {K.}~\bibnamefont {Gibble}},\ }\bibfield  {title} {\bibinfo
  {title} {Advances in the accuracy, stability, and reliability of the {PTB}
  primary fountain clocks},\ }\href {https://doi.org/10.1088/1681-7575/aae008}
  {\bibfield  {journal} {\bibinfo  {journal} {Metrologia}\ }\textbf {\bibinfo
  {volume} {55}},\ \bibinfo {pages} {789} (\bibinfo {year} {2018})}\BibitemShut
  {NoStop}%
\bibitem [{\citenamefont {Deutsch}\ \emph {et~al.}(2010)\citenamefont
  {Deutsch}, \citenamefont {Ramirez-Martinez}, \citenamefont {Lacro{\^{u}}te},
  \citenamefont {Reinhard}, \citenamefont {Schneider}, \citenamefont {Fuchs},
  \citenamefont {Pi{\'{e}}chon}, \citenamefont {Lalo{\"e}}, \citenamefont
  {Reichel},\ and\ \citenamefont {Rosenbusch}}]{Deutsch2010}%
  \BibitemOpen
  \bibfield  {author} {\bibinfo {author} {\bibfnamefont {C.}~\bibnamefont
  {Deutsch}}, \bibinfo {author} {\bibfnamefont {F.}~\bibnamefont
  {Ramirez-Martinez}}, \bibinfo {author} {\bibfnamefont {C.}~\bibnamefont
  {Lacro{\^{u}}te}}, \bibinfo {author} {\bibfnamefont {F.}~\bibnamefont
  {Reinhard}}, \bibinfo {author} {\bibfnamefont {T.}~\bibnamefont {Schneider}},
  \bibinfo {author} {\bibfnamefont {J.~N.}\ \bibnamefont {Fuchs}}, \bibinfo
  {author} {\bibfnamefont {F.}~\bibnamefont {Pi{\'{e}}chon}}, \bibinfo {author}
  {\bibfnamefont {F.}~\bibnamefont {Lalo{\"e}}}, \bibinfo {author}
  {\bibfnamefont {J.}~\bibnamefont {Reichel}},\ and\ \bibinfo {author}
  {\bibfnamefont {P.}~\bibnamefont {Rosenbusch}},\ }\bibfield  {title}
  {\bibinfo {title} {Spin self-rephasing and very long coherence times in a
  trapped atomic ensemble},\ }\href
  {https://doi.org/10.1103/physrevlett.105.020401} {\bibfield  {journal}
  {\bibinfo  {journal} {Physical Review Letters}\ }\textbf {\bibinfo {volume}
  {105}},\ \bibinfo {pages} {020401} (\bibinfo {year} {2010})}\BibitemShut
  {NoStop}%
\bibitem [{\citenamefont {Martin}\ \emph {et~al.}(2013)\citenamefont {Martin},
  \citenamefont {Bishof}, \citenamefont {Swallows}, \citenamefont {Zhang},
  \citenamefont {Benko}, \citenamefont {von Stecher}, \citenamefont {Gorshkov},
  \citenamefont {Rey},\ and\ \citenamefont {Ye}}]{Martin2013}%
  \BibitemOpen
  \bibfield  {author} {\bibinfo {author} {\bibfnamefont {M.~J.}\ \bibnamefont
  {Martin}}, \bibinfo {author} {\bibfnamefont {M.}~\bibnamefont {Bishof}},
  \bibinfo {author} {\bibfnamefont {M.~D.}\ \bibnamefont {Swallows}}, \bibinfo
  {author} {\bibfnamefont {X.}~\bibnamefont {Zhang}}, \bibinfo {author}
  {\bibfnamefont {C.}~\bibnamefont {Benko}}, \bibinfo {author} {\bibfnamefont
  {J.}~\bibnamefont {von Stecher}}, \bibinfo {author} {\bibfnamefont {A.~V.}\
  \bibnamefont {Gorshkov}}, \bibinfo {author} {\bibfnamefont {A.~M.}\
  \bibnamefont {Rey}},\ and\ \bibinfo {author} {\bibfnamefont {J.}~\bibnamefont
  {Ye}},\ }\bibfield  {title} {\bibinfo {title} {A quantum many-body spin
  system in an optical lattice clock},\ }\href
  {https://doi.org/10.1126/science.1236929} {\bibfield  {journal} {\bibinfo
  {journal} {Science}\ }\textbf {\bibinfo {volume} {341}},\ \bibinfo {pages}
  {632} (\bibinfo {year} {2013})}\BibitemShut {NoStop}%
\bibitem [{\citenamefont {He}\ \emph {et~al.}(2019)\citenamefont {He},
  \citenamefont {Perlin}, \citenamefont {Muleady}, \citenamefont {Lewis-Swan},
  \citenamefont {Hutson}, \citenamefont {Ye},\ and\ \citenamefont
  {Rey}}]{He2019}%
  \BibitemOpen
  \bibfield  {author} {\bibinfo {author} {\bibfnamefont {P.}~\bibnamefont
  {He}}, \bibinfo {author} {\bibfnamefont {M.~A.}\ \bibnamefont {Perlin}},
  \bibinfo {author} {\bibfnamefont {S.~R.}\ \bibnamefont {Muleady}}, \bibinfo
  {author} {\bibfnamefont {R.~J.}\ \bibnamefont {Lewis-Swan}}, \bibinfo
  {author} {\bibfnamefont {R.~B.}\ \bibnamefont {Hutson}}, \bibinfo {author}
  {\bibfnamefont {J.}~\bibnamefont {Ye}},\ and\ \bibinfo {author}
  {\bibfnamefont {A.~M.}\ \bibnamefont {Rey}},\ }\bibfield  {title} {\bibinfo
  {title} {Engineering spin squeezing in a 3d optical lattice with interacting
  spin-orbit-coupled fermions},\ }\href
  {https://doi.org/10.1103/PhysRevResearch.1.033075} {\bibfield  {journal}
  {\bibinfo  {journal} {Phys. Rev. Research}\ }\textbf {\bibinfo {volume}
  {1}},\ \bibinfo {pages} {033075} (\bibinfo {year} {2019})}\BibitemShut
  {NoStop}%
\bibitem [{\citenamefont {Bilitewski}\ \emph {et~al.}(2021)\citenamefont
  {Bilitewski}, \citenamefont {De~Marco}, \citenamefont {Li}, \citenamefont
  {Matsuda}, \citenamefont {Tobias}, \citenamefont {Valtolina}, \citenamefont
  {Ye},\ and\ \citenamefont {Rey}}]{bilitewski_dynamical_2021}%
  \BibitemOpen
  \bibfield  {author} {\bibinfo {author} {\bibfnamefont {T.}~\bibnamefont
  {Bilitewski}}, \bibinfo {author} {\bibfnamefont {L.}~\bibnamefont
  {De~Marco}}, \bibinfo {author} {\bibfnamefont {J.-R.}\ \bibnamefont {Li}},
  \bibinfo {author} {\bibfnamefont {K.}~\bibnamefont {Matsuda}}, \bibinfo
  {author} {\bibfnamefont {W.~G.}\ \bibnamefont {Tobias}}, \bibinfo {author}
  {\bibfnamefont {G.}~\bibnamefont {Valtolina}}, \bibinfo {author}
  {\bibfnamefont {J.}~\bibnamefont {Ye}},\ and\ \bibinfo {author}
  {\bibfnamefont {A.~M.}\ \bibnamefont {Rey}},\ }\bibfield  {title} {\bibinfo
  {title} {Dynamical {Generation} of {Spin} {Squeezing} in {Ultracold}
  {Dipolar} {Molecules}},\ }\href
  {https://doi.org/10.1103/PhysRevLett.126.113401} {\bibfield  {journal}
  {\bibinfo  {journal} {Physical Review Letters}\ }\textbf {\bibinfo {volume}
  {126}},\ \bibinfo {pages} {113401} (\bibinfo {year} {2021})}\BibitemShut
  {NoStop}%
\bibitem [{\citenamefont {Szmuk}\ \emph {et~al.}(2015)\citenamefont {Szmuk},
  \citenamefont {Dugrain}, \citenamefont {Maineult}, \citenamefont {Reichel},\
  and\ \citenamefont {Rosenbusch}}]{Szmuk2015}%
  \BibitemOpen
  \bibfield  {author} {\bibinfo {author} {\bibfnamefont {R.}~\bibnamefont
  {Szmuk}}, \bibinfo {author} {\bibfnamefont {V.}~\bibnamefont {Dugrain}},
  \bibinfo {author} {\bibfnamefont {W.}~\bibnamefont {Maineult}}, \bibinfo
  {author} {\bibfnamefont {J.}~\bibnamefont {Reichel}},\ and\ \bibinfo {author}
  {\bibfnamefont {P.}~\bibnamefont {Rosenbusch}},\ }\bibfield  {title}
  {\bibinfo {title} {Stability of a trapped-atom clock on a chip},\ }\href
  {https://doi.org/10.1103/physreva.92.012106} {\bibfield  {journal} {\bibinfo
  {journal} {Physical Review A}\ }\textbf {\bibinfo {volume} {92}},\ \bibinfo
  {pages} {012106} (\bibinfo {year} {2015})}\BibitemShut {NoStop}%
\bibitem [{\citenamefont {Ott}\ \emph {et~al.}(2016)\citenamefont {Ott},
  \citenamefont {Garcia}, \citenamefont {Kohlhaas}, \citenamefont
  {Sch{\"u}ppert}, \citenamefont {Rosenbusch}, \citenamefont {Long},\ and\
  \citenamefont {Reichel}}]{Ott2016}%
  \BibitemOpen
  \bibfield  {author} {\bibinfo {author} {\bibfnamefont {K.}~\bibnamefont
  {Ott}}, \bibinfo {author} {\bibfnamefont {S.}~\bibnamefont {Garcia}},
  \bibinfo {author} {\bibfnamefont {R.}~\bibnamefont {Kohlhaas}}, \bibinfo
  {author} {\bibfnamefont {K.}~\bibnamefont {Sch{\"u}ppert}}, \bibinfo {author}
  {\bibfnamefont {P.}~\bibnamefont {Rosenbusch}}, \bibinfo {author}
  {\bibfnamefont {R.}~\bibnamefont {Long}},\ and\ \bibinfo {author}
  {\bibfnamefont {J.}~\bibnamefont {Reichel}},\ }\bibfield  {title} {\bibinfo
  {title} {Millimeter-long fiber fabry-perot cavities},\ }\href
  {https://doi.org/10.1364/oe.24.009839} {\bibfield  {journal} {\bibinfo
  {journal} {Optics Express}\ }\textbf {\bibinfo {volume} {24}},\ \bibinfo
  {pages} {9839} (\bibinfo {year} {2016})}\BibitemShut {NoStop}%
\bibitem [{\citenamefont {Harber}\ \emph {et~al.}(2002)\citenamefont {Harber},
  \citenamefont {Lewandowski}, \citenamefont {McGuirk},\ and\ \citenamefont
  {Cornell}}]{Harber2002}%
  \BibitemOpen
  \bibfield  {author} {\bibinfo {author} {\bibfnamefont {D.~M.}\ \bibnamefont
  {Harber}}, \bibinfo {author} {\bibfnamefont {H.~J.}\ \bibnamefont
  {Lewandowski}}, \bibinfo {author} {\bibfnamefont {J.~M.}\ \bibnamefont
  {McGuirk}},\ and\ \bibinfo {author} {\bibfnamefont {E.~A.}\ \bibnamefont
  {Cornell}},\ }\bibfield  {title} {\bibinfo {title} {Effect of cold collisions
  on spin coherence and resonance shifts in a magnetically trapped ultracold
  gas},\ }\href {https://doi.org/10.1103/PhysRevA.66.053616} {\bibfield
  {journal} {\bibinfo  {journal} {Physical Review A}\ }\textbf {\bibinfo
  {volume} {66}},\ \bibinfo {pages} {053616} (\bibinfo {year}
  {2002})}\BibitemShut {NoStop}%
\bibitem [{\citenamefont {Li}\ \emph {et~al.}(2008)\citenamefont {Li},
  \citenamefont {Castin},\ and\ \citenamefont {Sinatra}}]{Li2008}%
  \BibitemOpen
  \bibfield  {author} {\bibinfo {author} {\bibfnamefont {Y.}~\bibnamefont
  {Li}}, \bibinfo {author} {\bibfnamefont {Y.}~\bibnamefont {Castin}},\ and\
  \bibinfo {author} {\bibfnamefont {A.}~\bibnamefont {Sinatra}},\ }\bibfield
  {title} {\bibinfo {title} {Optimum spin squeezing in bose-einstein
  condensates with particle losses},\ }\href
  {https://doi.org/10.1103/physrevlett.100.210401} {\bibfield  {journal}
  {\bibinfo  {journal} {Physical Review Letters}\ }\textbf {\bibinfo {volume}
  {100}},\ \bibinfo {pages} {210401} (\bibinfo {year} {2008})}\BibitemShut
  {NoStop}%
\bibitem [{\citenamefont {Kleine~B\"uning}\ \emph {et~al.}(2011)\citenamefont
  {Kleine~B\"uning}, \citenamefont {Will}, \citenamefont {Ertmer},
  \citenamefont {Rasel}, \citenamefont {Arlt}, \citenamefont {Klempt},
  \citenamefont {Ramirez-Martinez}, \citenamefont {Pi\'echon},\ and\
  \citenamefont {Rosenbusch}}]{Kleine2011}%
  \BibitemOpen
  \bibfield  {author} {\bibinfo {author} {\bibfnamefont {G.}~\bibnamefont
  {Kleine~B\"uning}}, \bibinfo {author} {\bibfnamefont {J.}~\bibnamefont
  {Will}}, \bibinfo {author} {\bibfnamefont {W.}~\bibnamefont {Ertmer}},
  \bibinfo {author} {\bibfnamefont {E.}~\bibnamefont {Rasel}}, \bibinfo
  {author} {\bibfnamefont {J.}~\bibnamefont {Arlt}}, \bibinfo {author}
  {\bibfnamefont {C.}~\bibnamefont {Klempt}}, \bibinfo {author} {\bibfnamefont
  {F.}~\bibnamefont {Ramirez-Martinez}}, \bibinfo {author} {\bibfnamefont
  {F.}~\bibnamefont {Pi\'echon}},\ and\ \bibinfo {author} {\bibfnamefont
  {P.}~\bibnamefont {Rosenbusch}},\ }\bibfield  {title} {\bibinfo {title}
  {Extended coherence time on the clock transition of optically trapped
  rubidium},\ }\href {https://doi.org/10.1103/PhysRevLett.106.240801}
  {\bibfield  {journal} {\bibinfo  {journal} {Phys. Rev. Lett.}\ }\textbf
  {\bibinfo {volume} {106}},\ \bibinfo {pages} {240801} (\bibinfo {year}
  {2011})}\BibitemShut {NoStop}%
\bibitem [{\citenamefont {Solaro}\ \emph {et~al.}(2016)\citenamefont {Solaro},
  \citenamefont {Bonnin}, \citenamefont {Combes}, \citenamefont {Lopez},
  \citenamefont {Alauze}, \citenamefont {Fuchs}, \citenamefont
  {Pi{\'{e}}chon},\ and\ \citenamefont {Santos}}]{Solaro2016}%
  \BibitemOpen
  \bibfield  {author} {\bibinfo {author} {\bibfnamefont {C.}~\bibnamefont
  {Solaro}}, \bibinfo {author} {\bibfnamefont {A.}~\bibnamefont {Bonnin}},
  \bibinfo {author} {\bibfnamefont {F.}~\bibnamefont {Combes}}, \bibinfo
  {author} {\bibfnamefont {M.}~\bibnamefont {Lopez}}, \bibinfo {author}
  {\bibfnamefont {X.}~\bibnamefont {Alauze}}, \bibinfo {author} {\bibfnamefont
  {J.-N.}\ \bibnamefont {Fuchs}}, \bibinfo {author} {\bibfnamefont
  {F.}~\bibnamefont {Pi{\'{e}}chon}},\ and\ \bibinfo {author} {\bibfnamefont
  {F.~P.~D.}\ \bibnamefont {Santos}},\ }\bibfield  {title} {\bibinfo {title}
  {Competition between spin echo and spin self-rephasing in a trapped atom
  interferometer},\ }\href {https://doi.org/10.1103/physrevlett.117.163003}
  {\bibfield  {journal} {\bibinfo  {journal} {Physical Review Letters}\
  }\textbf {\bibinfo {volume} {117}},\ \bibinfo {pages} {163003} (\bibinfo
  {year} {2016})}\BibitemShut {NoStop}%
\bibitem [{\citenamefont {Hu}\ \emph {et~al.}(2015)\citenamefont {Hu},
  \citenamefont {Chen}, \citenamefont {Vendeiro}, \citenamefont {Zhang},\ and\
  \citenamefont {Vuleti{\'{c}}}}]{Hu2015}%
  \BibitemOpen
  \bibfield  {author} {\bibinfo {author} {\bibfnamefont {J.}~\bibnamefont
  {Hu}}, \bibinfo {author} {\bibfnamefont {W.}~\bibnamefont {Chen}}, \bibinfo
  {author} {\bibfnamefont {Z.}~\bibnamefont {Vendeiro}}, \bibinfo {author}
  {\bibfnamefont {H.}~\bibnamefont {Zhang}},\ and\ \bibinfo {author}
  {\bibfnamefont {V.}~\bibnamefont {Vuleti{\'{c}}}},\ }\bibfield  {title}
  {\bibinfo {title} {Entangled collective-spin states of atomic ensembles under
  nonuniform atom-light interaction},\ }\href
  {https://doi.org/10.1103/physreva.92.063816} {\bibfield  {journal} {\bibinfo
  {journal} {Physical Review A}\ }\textbf {\bibinfo {volume} {92}},\ \bibinfo
  {pages} {063816} (\bibinfo {year} {2015})}\BibitemShut {NoStop}%
\bibitem [{\citenamefont {Davis}\ \emph {et~al.}(2016)\citenamefont {Davis},
  \citenamefont {Bentsen},\ and\ \citenamefont {Schleier-Smith}}]{Davis2016}%
  \BibitemOpen
  \bibfield  {author} {\bibinfo {author} {\bibfnamefont {E.}~\bibnamefont
  {Davis}}, \bibinfo {author} {\bibfnamefont {G.}~\bibnamefont {Bentsen}},\
  and\ \bibinfo {author} {\bibfnamefont {M.}~\bibnamefont {Schleier-Smith}},\
  }\bibfield  {title} {\bibinfo {title} {Approaching the heisenberg limit
  without single-particle detection},\ }\href
  {https://doi.org/10.1103/physrevlett.116.053601} {\bibfield  {journal}
  {\bibinfo  {journal} {Physical Review Letters}\ }\textbf {\bibinfo {volume}
  {116}},\ \bibinfo {pages} {053601} (\bibinfo {year} {2016})}\BibitemShut
  {NoStop}%
\bibitem [{\citenamefont {Hosten}\ \emph
  {et~al.}(2016{\natexlab{b}})\citenamefont {Hosten}, \citenamefont
  {Krishnakumar}, \citenamefont {Engelsen},\ and\ \citenamefont
  {Kasevich}}]{Hosten2016s}%
  \BibitemOpen
  \bibfield  {author} {\bibinfo {author} {\bibfnamefont {O.}~\bibnamefont
  {Hosten}}, \bibinfo {author} {\bibfnamefont {R.}~\bibnamefont
  {Krishnakumar}}, \bibinfo {author} {\bibfnamefont {N.~J.}\ \bibnamefont
  {Engelsen}},\ and\ \bibinfo {author} {\bibfnamefont {M.~A.}\ \bibnamefont
  {Kasevich}},\ }\bibfield  {title} {\bibinfo {title} {Quantum phase
  magnification},\ }\href {https://doi.org/10.1126/science.aaf3397} {\bibfield
  {journal} {\bibinfo  {journal} {Science}\ }\textbf {\bibinfo {volume}
  {352}},\ \bibinfo {pages} {1552} (\bibinfo {year}
  {2016}{\natexlab{b}})}\BibitemShut {NoStop}%
\bibitem [{\citenamefont {Smale}\ \emph {et~al.}(2019)\citenamefont {Smale},
  \citenamefont {He}, \citenamefont {Olsen}, \citenamefont {Jackson},
  \citenamefont {Sharum}, \citenamefont {Trotzky}, \citenamefont {Marino},
  \citenamefont {Rey},\ and\ \citenamefont {Thywissen}}]{Smale2019}%
  \BibitemOpen
  \bibfield  {author} {\bibinfo {author} {\bibfnamefont {S.}~\bibnamefont
  {Smale}}, \bibinfo {author} {\bibfnamefont {P.}~\bibnamefont {He}}, \bibinfo
  {author} {\bibfnamefont {B.~A.}\ \bibnamefont {Olsen}}, \bibinfo {author}
  {\bibfnamefont {K.~G.}\ \bibnamefont {Jackson}}, \bibinfo {author}
  {\bibfnamefont {H.}~\bibnamefont {Sharum}}, \bibinfo {author} {\bibfnamefont
  {S.}~\bibnamefont {Trotzky}}, \bibinfo {author} {\bibfnamefont
  {J.}~\bibnamefont {Marino}}, \bibinfo {author} {\bibfnamefont {A.~M.}\
  \bibnamefont {Rey}},\ and\ \bibinfo {author} {\bibfnamefont {J.~H.}\
  \bibnamefont {Thywissen}},\ }\bibfield  {title} {\bibinfo {title}
  {Observation of a transition between dynamical phases in a quantum degenerate
  fermi gas},\ }\href {https://doi.org/10.1126/sciadv.aax1568} {\bibfield
  {journal} {\bibinfo  {journal} {Science Advances}\ }\textbf {\bibinfo
  {volume} {5}},\ \bibinfo {pages} {eaax1568} (\bibinfo {year}
  {2019})}\BibitemShut {NoStop}%
\bibitem [{\citenamefont {Norcia}\ \emph {et~al.}(2018)\citenamefont {Norcia},
  \citenamefont {Lewis-Swan}, \citenamefont {Cline}, \citenamefont {Zhu},
  \citenamefont {Rey},\ and\ \citenamefont {Thompson}}]{Norcia2018b}%
  \BibitemOpen
  \bibfield  {author} {\bibinfo {author} {\bibfnamefont {M.~A.}\ \bibnamefont
  {Norcia}}, \bibinfo {author} {\bibfnamefont {R.~J.}\ \bibnamefont
  {Lewis-Swan}}, \bibinfo {author} {\bibfnamefont {J.~R.~K.}\ \bibnamefont
  {Cline}}, \bibinfo {author} {\bibfnamefont {B.}~\bibnamefont {Zhu}}, \bibinfo
  {author} {\bibfnamefont {A.~M.}\ \bibnamefont {Rey}},\ and\ \bibinfo {author}
  {\bibfnamefont {J.~K.}\ \bibnamefont {Thompson}},\ }\bibfield  {title}
  {\bibinfo {title} {Cavity-mediated collective spin-exchange interactions in a
  strontium superradiant laser},\ }\href
  {https://doi.org/10.1126/science.aar3102} {\bibfield  {journal} {\bibinfo
  {journal} {Science}\ }\textbf {\bibinfo {volume} {361}},\ \bibinfo {pages}
  {259} (\bibinfo {year} {2018})}\BibitemShut {NoStop}%
\bibitem [{\citenamefont {Davis}\ \emph {et~al.}(2019)\citenamefont {Davis},
  \citenamefont {Bentsen}, \citenamefont {Homeier}, \citenamefont {Li},\ and\
  \citenamefont {Schleier-Smith}}]{Davis2019}%
  \BibitemOpen
  \bibfield  {author} {\bibinfo {author} {\bibfnamefont {E.~J.}\ \bibnamefont
  {Davis}}, \bibinfo {author} {\bibfnamefont {G.}~\bibnamefont {Bentsen}},
  \bibinfo {author} {\bibfnamefont {L.}~\bibnamefont {Homeier}}, \bibinfo
  {author} {\bibfnamefont {T.}~\bibnamefont {Li}},\ and\ \bibinfo {author}
  {\bibfnamefont {M.~H.}\ \bibnamefont {Schleier-Smith}},\ }\bibfield  {title}
  {\bibinfo {title} {Photon-mediated spin-exchange dynamics of spin-1 atoms},\
  }\href {https://doi.org/10.1103/physrevlett.122.010405} {\bibfield  {journal}
  {\bibinfo  {journal} {Physical Review Letters}\ }\textbf {\bibinfo {volume}
  {122}},\ \bibinfo {pages} {010405} (\bibinfo {year} {2019})}\BibitemShut
  {NoStop}%
\bibitem [{\citenamefont {Cummins}\ \emph {et~al.}(2003)\citenamefont
  {Cummins}, \citenamefont {Llewellyn},\ and\ \citenamefont
  {Jones}}]{Cummins2003}%
  \BibitemOpen
  \bibfield  {author} {\bibinfo {author} {\bibfnamefont {H.~K.}\ \bibnamefont
  {Cummins}}, \bibinfo {author} {\bibfnamefont {G.}~\bibnamefont {Llewellyn}},\
  and\ \bibinfo {author} {\bibfnamefont {J.~A.}\ \bibnamefont {Jones}},\
  }\bibfield  {title} {\bibinfo {title} {Tackling systematic errors in quantum
  logic gates with composite rotations},\ }\href
  {https://doi.org/10.1103/physreva.67.042308} {\bibfield  {journal} {\bibinfo
  {journal} {Physical Review A}\ }\textbf {\bibinfo {volume} {67}},\ \bibinfo
  {pages} {042308} (\bibinfo {year} {2003})}\BibitemShut {NoStop}%
\bibitem [{\citenamefont {Braverman}\ \emph {et~al.}(2018)\citenamefont
  {Braverman}, \citenamefont {Kawasaki},\ and\ \citenamefont
  {Vuleti{\'{c}}}}]{Braverman2018}%
  \BibitemOpen
  \bibfield  {author} {\bibinfo {author} {\bibfnamefont {B.}~\bibnamefont
  {Braverman}}, \bibinfo {author} {\bibfnamefont {A.}~\bibnamefont
  {Kawasaki}},\ and\ \bibinfo {author} {\bibfnamefont {V.}~\bibnamefont
  {Vuleti{\'{c}}}},\ }\bibfield  {title} {\bibinfo {title} {Impact of
  non-unitary spin squeezing on atomic clock performance},\ }\href
  {https://doi.org/10.1088/1367-2630/aae563} {\bibfield  {journal} {\bibinfo
  {journal} {New Journal of Physics}\ }\textbf {\bibinfo {volume} {20}},\
  \bibinfo {pages} {103019} (\bibinfo {year} {2018})}\BibitemShut {NoStop}%
\bibitem [{\citenamefont {Braverman}\ \emph {et~al.}(2019)\citenamefont
  {Braverman}, \citenamefont {Kawasaki}, \citenamefont
  {Pedrozo-Pe{\~{n}}afiel}, \citenamefont {Colombo}, \citenamefont {Shu},
  \citenamefont {Li}, \citenamefont {Mendez}, \citenamefont {Yamoah},
  \citenamefont {Salvi}, \citenamefont {Akamatsu}, \citenamefont {Xiao},\ and\
  \citenamefont {Vuleti{\'{c}}}}]{Braverman2019}%
  \BibitemOpen
  \bibfield  {author} {\bibinfo {author} {\bibfnamefont {B.}~\bibnamefont
  {Braverman}}, \bibinfo {author} {\bibfnamefont {A.}~\bibnamefont {Kawasaki}},
  \bibinfo {author} {\bibfnamefont {E.}~\bibnamefont {Pedrozo-Pe{\~{n}}afiel}},
  \bibinfo {author} {\bibfnamefont {S.}~\bibnamefont {Colombo}}, \bibinfo
  {author} {\bibfnamefont {C.}~\bibnamefont {Shu}}, \bibinfo {author}
  {\bibfnamefont {Z.}~\bibnamefont {Li}}, \bibinfo {author} {\bibfnamefont
  {E.}~\bibnamefont {Mendez}}, \bibinfo {author} {\bibfnamefont
  {M.}~\bibnamefont {Yamoah}}, \bibinfo {author} {\bibfnamefont
  {L.}~\bibnamefont {Salvi}}, \bibinfo {author} {\bibfnamefont
  {D.}~\bibnamefont {Akamatsu}}, \bibinfo {author} {\bibfnamefont
  {Y.}~\bibnamefont {Xiao}},\ and\ \bibinfo {author} {\bibfnamefont
  {V.}~\bibnamefont {Vuleti{\'{c}}}},\ }\bibfield  {title} {\bibinfo {title}
  {Near-unitary spin squeezing in {Yb171}},\ }\href
  {https://doi.org/10.1103/PhysRevLett.122.223203} {\bibfield  {journal}
  {\bibinfo  {journal} {Physical Review Letters}\ }\textbf {\bibinfo {volume}
  {122}},\ \bibinfo {pages} {223203} (\bibinfo {year} {2019})}\BibitemShut
  {NoStop}%
\bibitem [{\citenamefont {Pi{\'{e}}chon}\ \emph {et~al.}(2009)\citenamefont
  {Pi{\'{e}}chon}, \citenamefont {Fuchs},\ and\ \citenamefont
  {Lalo{\"e}}}]{Piechon2009}%
  \BibitemOpen
  \bibfield  {author} {\bibinfo {author} {\bibfnamefont {F.}~\bibnamefont
  {Pi{\'{e}}chon}}, \bibinfo {author} {\bibfnamefont {J.~N.}\ \bibnamefont
  {Fuchs}},\ and\ \bibinfo {author} {\bibfnamefont {F.}~\bibnamefont
  {Lalo{\"e}}},\ }\bibfield  {title} {\bibinfo {title} {Cumulative identical
  spin rotation effects in collisionless trapped atomic gases},\ }\href
  {https://doi.org/10.1103/physrevlett.102.215301} {\bibfield  {journal}
  {\bibinfo  {journal} {Physical Review Letters}\ }\textbf {\bibinfo {volume}
  {102}},\ \bibinfo {pages} {215301} (\bibinfo {year} {2009})}\BibitemShut
  {NoStop}%
\bibitem [{\citenamefont {Fuchs}\ \emph {et~al.}(2002)\citenamefont {Fuchs},
  \citenamefont {Gangardt},\ and\ \citenamefont {Lalo{\"e}}}]{Fuchs2002}%
  \BibitemOpen
  \bibfield  {author} {\bibinfo {author} {\bibfnamefont {J.~N.}\ \bibnamefont
  {Fuchs}}, \bibinfo {author} {\bibfnamefont {D.~M.}\ \bibnamefont
  {Gangardt}},\ and\ \bibinfo {author} {\bibfnamefont {F.}~\bibnamefont
  {Lalo{\"e}}},\ }\bibfield  {title} {\bibinfo {title} {Internal state
  conversion in ultracold gases},\ }\href
  {https://doi.org/10.1103/physrevlett.88.230404} {\bibfield  {journal}
  {\bibinfo  {journal} {Physical Review Letters}\ }\textbf {\bibinfo {volume}
  {88}},\ \bibinfo {pages} {230404} (\bibinfo {year} {2002})}\BibitemShut
  {NoStop}%
\bibitem [{\citenamefont {Pegahan}\ \emph {et~al.}(2018)\citenamefont
  {Pegahan}, \citenamefont {Kangara}, \citenamefont {Arakelyan},\ and\
  \citenamefont {Thomas}}]{Pegahan2018}%
  \BibitemOpen
  \bibfield  {author} {\bibinfo {author} {\bibfnamefont {S.}~\bibnamefont
  {Pegahan}}, \bibinfo {author} {\bibfnamefont {J.}~\bibnamefont {Kangara}},
  \bibinfo {author} {\bibfnamefont {I.}~\bibnamefont {Arakelyan}},\ and\
  \bibinfo {author} {\bibfnamefont {J.~E.}\ \bibnamefont {Thomas}},\ }\bibfield
   {title} {\bibinfo {title} {Spin-energy correlation in degenerate
  weakly-interacting fermi gases},\ }\href
  {https://doi.org/10.1103/PhysRevA.99.063620} {\bibfield  {journal} {\bibinfo
  {journal} {Physical Review A}\ }\textbf {\bibinfo {volume} {99}},\ \bibinfo
  {pages} {063620} (\bibinfo {year} {2018})}\BibitemShut {NoStop}%
\end{thebibliography}%

\end{document}